\documentclass{article}

\usepackage{arxiv}

\usepackage[utf8]{inputenc} 
\usepackage[T1]{fontenc}    
\usepackage{hyperref}       
\usepackage{url}            
\usepackage{booktabs}       
\usepackage{amsfonts}       
\usepackage{nicefrac}       
\usepackage{microtype}      
\usepackage{graphicx}
\usepackage{natbib}
\usepackage{doi}

\usepackage{graphicx}

\usepackage{algorithm}
\usepackage{algorithmic}

\newtheorem{theorem}{Theorem}

\newtheorem{definition}{Definition}

\usepackage{multirow}
\usepackage{mathrsfs} 
\usepackage{amsfonts} 
\usepackage{comment}
\usepackage{subfig}
\usepackage{appendix}

\newcommand{\ie}{{\it i.e.,~}}
\newcommand{\eg}{{\it e.g.,~}}
\newcommand{\etal}{{\it et al.}}

\newcommand{\CC}{\mathcal{C}}

\newcommand{\CA}{\mathscr{A}}
\newcommand{\CB}{\mathscr{B}}
\newcommand{\CS}{\mathbb{*}}

\newcommand{\XX}{\mathsf{X}}
\newcommand{\YY}{\mathsf{Y}}
\newcommand{\Co}{\mathsf{C}}

\newcommand{\yi}{\Upsilon}

\newcommand{\myC}{\mathsf{my}\mathcal{C}}
\newcommand{\LookCS}{\mathtt{LookCS()}}
\newcommand{\MyLocation}{\mathtt{MyLocation()}}

\newcommand{\BigO}{\mathcal{O}}

\newcommand{\FSYNC}{\mathsf{FSYNC}}
\newcommand{\SSYNC}{\mathsf{SSYNC}}
\newcommand{\ASYNC}{\mathsf{ASYNC}}

\title{Complete Visibility Algorithm for Autonomous Mobile Luminous Robots under an Asynchronous Scheduler on Grid Plane}

\author{ \href{https://orcid.org/0000-0002-5437-7626}{\includegraphics[scale=0.06]{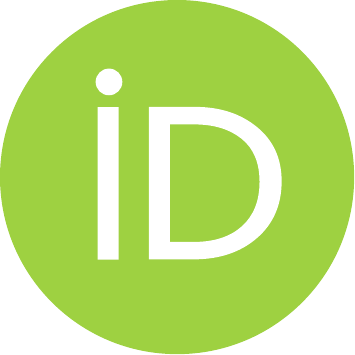}\hspace{1mm}Yonghwan Kim}\\
	Department of Computer Science\\
	Nagoya Institute of Technology\\
	Aichi, Japan \\
	\texttt{kim@nitech.ac.jp} \\
	\And
	\href{https://orcid.org/0000-0003-1683-2154}{\includegraphics[scale=0.06]{orcid.pdf}\hspace{1mm}Yoshiaki Katayama} \\
	Department of Computer Science\\
	Nagoya Institute of Technology\\
	Aichi, Japan \\
	\texttt{katayama@nitech.ac.jp} \\
	\AND
	\href{https://orcid.org/0000-0002-5351-1459}{\includegraphics[scale=0.06]{orcid.pdf}\hspace{1mm}Koichi Wada} \\
	Department of Applied Informatics\\
	Hosei University\\
	Tokyo, Japan \\
	\texttt{wada@hosei.ac.jp} \\
}

\date{}

\renewcommand{\shorttitle}{Complete Visibility by Luminous Robots on Grid}

\hypersetup{
pdftitle={A template for the arxiv style},
pdfsubject={q-bio.NC, q-bio.QM},
pdfauthor={David S.~Hippocampus, Elias D.~Striatum},
pdfkeywords={First keyword, Second keyword, More},
}

\begin{document}
\maketitle

\begin{abstract}
An autonomous mobile robot system is a distributed system consisting of mobile computational entities (called robots) that autonomously and repeatedly perform three operations: Look, Compute, and Move.
Various problems related to autonomous mobile robots, such as gathering, pattern formation, or flocking, have been extensively studied to understand the relationship between each robot's capabilities and the solvability of these problems.
In this study, we focus on the complete visibility problem, which involves relocating all the robots on an infinite grid plane such that each robot is visible to every other robot. We assume that each robot is a luminous robot 
(\ie has a light with a constant number of colors) and opaque (not transparent).
In this paper, we propose an algorithm to achieve  complete visibility when a set of robots is given. 
The algorithm ensures that complete visibility is achieved even when robots operate asynchronously and have no knowledge of the total number of robots on the grid plane using only two colors.
\end{abstract}


\section{Introduction}
\label{sec:typesetting-summary}
An autonomous mobile robot system is a distributed system composed of mobile computational entities known as \emph{robots}
which operate and move autonomously. 
Each robot determines its destination based solely on the (relative) locations of other robots
obtained from its own (\ie local) observation,
and cooperate with  others to achieve a common goal for the system.
The fundamental model is proposed by Suzuki \etal, wherein  
the robots possess no communication abilities (\emph{silent}), 
have no memory to recall 
 past activities (\emph{oblivious}), and 
lack identifiers (\emph{anonymous}).
Furthermore, all robots are indistinguishable from each other (\emph{identical}) and execute the same algorithm (\emph{homogeneous}).
Each robot continuously executes a cycle consisting of three phases; \textsl{Look, Compute} and \textsl{Move}. In the Look phase, a robot observes the positions of other robots. In the Compute phase, a robot caluculates its destination using an algorithm based on the results of the last Look phase, then proceeds to the calculated  destination during the Move phase.
In general, the level of synchronization between phase executions for each robot differs, leading to three basic schedulers: fully-synchronous ($\FSYNC$), semi-synchronous ($\SSYNC$), and asynchronous ($\ASYNC$).
In an $\FSYNC$ scheduler model, all robots perform the cycle in a fully synchronous manner,
meaning all robots execute the same phase synchronously.
$\SSYNC$ is a scheduler model that allows some robots to not execute a cycle while all robots are executing a cycle in $\FSYNC$.
$\ASYNC$ is a scheduler model that makes no assumptions about synchronization between robots' executions.

Several problems are considered in autonomous mobile robots systems under various assumptions (\ie capabilities of the robots), 
such as gathering \cite{gathering0},  deployment  \cite{distribute3}, and pattern formation problem \cite{Flo99}.
The solvability of  a problem is sensitively changed by the capability of the robots; 
therefore, researchers primarily aim to define 
 the minimum capability required to solve a  given problem.
Many autonomous mobile robot models neither have an explicit communication function nor memory.
However, a new model, termed the  \emph{luminous model}, has been introduced \cite{Luna2}.
A luminous robot is equipped with a color-changing light 
that can be observed by itself and/or other robots, 
and can communicate with other robots and remember its own status using its light. 
The luminous model clearly provides greater functionality than the basic model.
Indeed, research has demonstrated that  problems previously unsolvable using the basic model can addressed effectively  with the luminous model \cite{Luna2,Sharma2}.
In luminous robot systems, a primary area of interest for researchers is the number of light colors needed to solve a problem. 


Observability among  robots to each other is a critical aspect in these systems.
If a robot is opaque, it can only observe its immediate neighbors  when three or more robots are align in a straight line (collinear). Consequently, it becomes 
 impossible to recognize the configuration of the entire system based on personal observation alone. 
To address this issue, the complete visibility problem has been studied extensively \cite{Luna2,Luna0,Luna1,Sharma1,Sharma2,Sharma5,Sharma6,Sharma3,Sharma4}.
The complete visibility problem is a situation where all robots are required to locate themselves at points where they can mutually observe each other in a Euclidean or grid plane.
According to known results of this problem on a grid plane \cite{Adhikary,Sharma5,Sharma6}, more than 10 colors of light are required when there is no agreement on the coordinate system.
As a preliminary step in  investigating the trade-off between the degree  of agreement on the coordinate system and the number of colors of lights required to solve the problem, we propose an algorithm that addresses the problem using only two colors of light within a system that has agreement on the orientation and direction of grid plane's two axes.
This finding suggests   that by establishing consensus on the coordinate system, the number of required colors can be drastically reduced. 


\begin{table*}[b]
 \centering
    \begin{tabular}{cccccc}
    \hline 
   {\bf Work} & {\bf \# of colors} & {\bf Time Complexity} & {\bf Area} 
   & {\bf Agreement} & {\bf Algorithm} \\
   \hline 
   \cite{Adhikary} & 11 & finite$^\S$ & finite$^\S$ &  none & deterministic \\ 
   \cite{Sharma5} & 50 & $\mathcal{O} (D+n)$ \emph{w.h.p.} & $\mathcal{O} (n^2)$ & none & randomized \\
   \cite{Sharma5} & 17 & $\mathcal{O} (D+n)$ & $\mathcal{O} (n^2)$ & none & deterministic$^\dag$ \\ 
   \cite{Sharma6} & 32 & $\mathcal{O} (D+n)$ \emph{w.h.p.} & $\mathcal{O} (n^2)$ & none & randomized \\
   This work & 2 & $\mathcal{O} ((D+n))$ & $\mathcal{O} (n^2)$ & $X$ and $Y$ axes & deterministic \\
   \hline 
   & \multicolumn{5}{l}{\bf Lower bounds of time and spatial (area) complexities}\\
   \hline
   \cite{Sharma5,Sharma6} & any & $\Omega(n)$ & any & any & any \\
   \cite{FL92} & any & any & $\Omega(n^2)$ & any & any \\
   \hline
    \multicolumn{6}{l}{\small $^\S$The complexities are not provided in \cite{Adhikary}. It is shown in \cite{Sharma6} that the algorithm}\\
    \multicolumn{6}{l}{\small \ \ in \cite{Adhikary} requires $\Omega(Dn+n^2)$ time and $\Omega(n^2)$ area.}\\
   \multicolumn{6}{l}{\small $^\dag$The algorithm operates only when the initial configuration is not symmetric.}\\
   & & & & 
  \end{tabular}
   \caption{Comparison table between the previous works and this work and the known lower bounds of the complexities}
 \label{comp}
\end{table*}

\noindent
{\bf Related Works}:
The complete visibility is firstly studied by G. Di Luna \etal~\cite{Luna0},
which shows that a set of robots with a constant-sized color of lights (\ie luminous model) 
can solve the complete visibility problem on an Euclidean plane under $\ASYNC$, 
even there is no assumption on the number of robots, rigidity of movements, nor chirality.
They also proposed an algorithm for oblivious robots 
under $\SSYNC$ in \cite{Luna1} .
Moreover, G. Di Luna \etal~investigate under what conditions luminous robots 
can solve the complete visibility problem, and at what cost, in terms of the number of colors used by the robots 
in \cite{Luna2}.
They provided a spectrum of results which depends on the scheduler, 
on the number of robots, and on the number of colors.
Two algorithms, Shrink and Contain algorithms, are introduced in \cite{Luna2}, 
however, the time analyses of both algorithms are not provided.
G. Sharma \etal~analyzed and proved these two algorithms:
they showed that Shrink algorithm required $\Theta$($n$) 
and the time complexity of Contain algorithm is $\Omega$($n^2$)
, where $n$ is the total number of robots, in \cite{Sharma1}.
Furthermore, Sharma \etal~proposed an improved version of Shrink algorithm 
(called Modified Shrink algorithm) which decreases its time complexity to $\mathcal{O}$($n\log n$).
The optimal number of colors to solve the complete visibility problem is 
discussed in \cite{Sharma2};
this paper presents the optimal number of colors 
when a scheduler is assumed as $\SSYNC$ or $\ASYNC$, 
and/or robots are non-rigid or rigid.
G. Sharma \etal~proposed the first algorithm for complete visibility with 
$\mathcal{O}$(1) runtime under $\SSYNC$ in \cite{Sharma3}, 
and $\mathcal{O}$($\log n$)-time complete visibility algorithm 
for asynchronous robots with lights.


R. Adhikary \etal~\cite{Adhikary} were the  first to investigate the complete visibility problem on a  grid plane,
introducing a deterministic and asynchronous algorithm under the assumptions that 
robots do not agree on the global coordinates nor chirality, and
they do not know the total number of robots, using 11 colors of lights.
However, neither time nor spatial analysis of the algorithm was provided in \cite{Adhikary} 
(only the finite-time termination is stated).
These complexities were later addressed in  another paper \cite{Sharma6} after several years, which  
 shows that the time complexity of this algorithm is $\Omega$($Dn+n^2$)
and the spatial complexity 
\footnote{The spatial complexity represents the area of the smallest rectangle including all robots 
when the algorithm terminates (\ie complete visibility is achieved).}
is $\Omega$($n^2$),  
where $n$ is the total number of robots and $D$ is the diameter in the initial configuration.

G. Sharma \etal~proposed  randomized complete visibility algorithms using 50 colors of lights in \cite{Sharma5},
and introduced another deterministic algorithm using 17 colors of light 
which ensures the achievement of complete visibility only when an initial configuration is non-symmetric.
They reduced the number of colors from 50 to 32 for the randomized algorithm in \cite{Sharma6}.
These algorithms solves the complete visibility problem in $\mathcal{O}$($D+n$)-time and 
$\mathcal{O}$($n^2$)-space which are optimal;
the lower bounds of the time complexity and spatial complexity are also provided in \cite{Sharma6}.


In this study, we focus on reducing  the required number of colors 
for the complete visibility problem on a grid plane, 
even under stronger assumptions.
From another perspective,   
we are interested in using the required number of colors as a measure 
of the changes in computational power of robots under different assumptions.
As a result, we have  drastically reduced the number of colors to two, 
whereas all the previous works \cite{Adhikary,Sharma5,Sharma6} on a grid plane 
require more than 10 colors,
if the robots share a common sense of directions (\ie agree on  both X and Y axes). 
Table~\ref{comp} shows the difference between the previous works \cite{Adhikary,Sharma5,Sharma6} and this work,
where $n$ is the number of robots, $D$ is the diameter of the initial configuration,
and the known lower bounds of time and spatial complexities.
Our proposed algorithm asynchronously solves the same problem 
using a smaller number of colors 
in optimal time and space when $D = \mathcal{O}$($n$) 
provided the robots agree on two axes.




\section{Model and Problem Definition}
\subsection{Model}

{\bf Robots and Grid Plane}:
Let $\mathcal{R}$=\{$r_1,r_2,...,r_n$\} be the set of autonomous mobile robots with a total number of robots $n$.
These robots have no identifiers and cannot be distinguished from each other by appearance.
Also, robots do not have any memory to maintain their own past executions, \ie oblivious, and do not know the total number of robots $n$.
The robots exist on infinite grid plane $\mathcal{P}$, and every point in $\mathcal{P}$ can be occupied by at most one robot. The robots agree on the directions and the orientations of both axes, X-axis and Y-axis, but the do not know the origin of $\mathcal{P}$, \ie they do not know their global coordinates. 
In this paper, without loss of generality, 
we assume that the positive direction of X-axis (resp. Y-axis) is \emph{right} (resp. \emph{up}).

\noindent
{\bf Movement of Robots}:
A robot can move to only one of the four grid points adjacent to the grid point where it currently exists.
Hence we assume that every robot moves asynchronously but instantaneously: each robot cannot be observed by any other robots while it is moving. 
This property is known as \emph{Move-Atomic} introduced in \cite{Flo99}. 

\noindent
{\bf Lights}:
Each robot maintains a constant-sized visible memory register called light that can be observed by other robots \cite{Luna2}.
Each light can be set by one among the constant number of colors.
The color of the light can be changed at the end of Compute in the LCM cycle (will be described later) if necessary.
In other words, the robot decides where to move and the color of the light based on the observation results, and starts moving after changing the color of the light.
In the proposed algorithm, we assume that a robot uses only two colors, $\CA$ and $\CB$, $\CC \in \{\CA,\CB\}$.

\noindent
{\bf Visibility Range and Transparency}:
We assume that there is no limit to the robot's field of view, 
\ie a robot has unlimited visibility range.
Each robot is not transparent, \ie opaque;
if robot $r_k$ exists on the line between two robots $r_i$ and $r_j$, 
then robots $r_i$ and $r_j$ cannot observe each other.

\noindent
{\bf Operations and Scheduler}:
Each robot performs one of the four operations: {\sl Look, Compute, Move}, and {\sl Wait}.
Let {\sl Look, Compute, Move} be one cycle (called LCM cycle).
{\bf (Wait)} A robot is in a standby state. A robot which is in Wait should perform its LCM operations again within the finite time. 
At the start, all robots are in the {\sl Wait} state.
{\bf (Look)} A robot gets the color of its own light. 
A robot gets the coordinates and light color of the other robots in its local coordinate system which is self-centric, 
\ie the origin is the position of the robot itself. As a result of {\sl Look}, a robot gets a set $\mathcal{V}_r=\{r_0,r_1,\cdots,r_M\}$ 
of the coordinates and light colors of the observed robots based on its local coordinate system.
{\bf (Compute)} A robot executes an algorithm to determine the color of its light and the destination point. 
As the result of the executing algorithm, a robot immediately changes the color of its light (if necessary).
{\bf (Move)} A robot instantaneously moves to the destination point determined by {\sl Compute} operation.

We consider an asynchronous scheduler ($\ASYNC$) which has no assumption on the timing of each robot's operation. 
This means that all robots perform their operations at unpredictable time instants and durations.

\noindent
{\bf Configuration}:
Let configuration $C_t$ be the set of the (global) coordinates and 
colors of all robots at a given time $t$: 
$C_t = \{(r_{1.x}^t,r_{1.y}^t,r_{1.\CC}^t), (r_{2.x}^t,r_{2.y}^t,r_{2.\CC}^t), \cdots$ $(r_{n.x}^t,r_{n.y}^t,r_{n.\CC}^t)\}$, 
where $r_{i.x}^t$ (resp. $r_{i.y}^t$) is the x-coordinate
(y-coordinate) of robot $r_i$ on grid $\mathcal{P}$, and
$r_{i.\CC}^t$ is the color of $r_i$'s light at time $t$.
Remind that no robot knows its global coordinate.

Configuration $C_t$ is changed into another configuration $C_{t+1}$ 
when at least one of the robots moves or changes the color of its light.
In other words, under an $\ASYNC$ scheduler, 
configuration $C_{t+1}$ is obtained by 
any atomic operation of any robot's operates {\sl Compute} or 
{\sl Move} at configuration $C_t$ (if it moves or changes its color).

\subsection{Complete Visibility Problem}
Here we define the complete visibility problem as the following.

\begin{definition}
{\bf The Complete Visibility Problem.}
Given a set of robots which is initially located on the distinct points in grid plane.
Algorithm $\mathcal{A}$ solves the complete visibility problem 
if $\mathcal{A}$ satisfies all the following conditions:
(1) Algorithm $\mathcal{A}$ eventually terminates; algorithm $\mathcal{A}$
eventually reaches a configuration such that no robot can move, 
and
(2) Every robot can observe all other robots, \ie any three robots are not collinear, 
when algorithm $\mathcal{A}$ terminates.
\end{definition}

\section{Proposed Algorithm}
\subsection{Basic Strategy}
In this subsection, we briefly give the basic strategy of the proposed algorithm for solving the complete visibility problem under $\ASYNC$. 
To achieve complete visibility, we refer to the previous study \cite{Roth}:
for any prime number $m$, it is known that 
any three distinct points $(i^2 \bmod m, i)$, for any distinct $i \leq m$, are not collinear.
Therefore, we aim to make a configuration such that all robots are located on 
$(i^2 \bmod m,i)$ for distinct $i$ and $n \leq m$, 
where $n$ is the total number of robots and $m$ is the smallest prime number which 
is equal to or greater than $n$.
The proposed algorithm to solve the complete visibility problem consists of the following 
three phases. 

\begin{enumerate}
    \item {\bf Line Formation Phase:} 
    The first phase of the proposed algorithm locates all robots on a straight line parallel to the Y-axis.
    \item {\bf Coordinate System Generation Phase:}
    In the second phase, the algorithm determines three reference points by moving three robots 
    to generate the agreed coordinate system, 
    because the robots do not agree on the global coordinate system, and no robot knows its global coordinates.
    All robots can recognize the current common coordinate system by referring these reference points. 
    Moreover, the robots which moves to the reference points can count the total number of robots $n$, and this information $n$ is (implicitly) propagated to all other robots by their positions.
    \item {\bf Complete Visibility Achievement Phase:}
    Each robot computes its destination point on the common coordinate system generated in the previous phase, 
    and moves to its appropriate position to achieve complete visibility.
\end{enumerate}

The proposed algorithm consists of 26 rules based on the observation result of each robot;
15 rules for the first phase, 8 rules of the second one, and 3 rules for the last phase.
Because of the lack of space, the detailed algorithm is presented in Appendix.

\subsection{Presentation of Algorithm}
Before the introduction of the proposed algorithm, 
we present some notations and functions
which are used in the presentation of the proposed algorithm.

We present the proposed algorithm using the conditions and action,
which means each robot executes the action when all conditions are satisfied, 
and represent as {\it $<$Conditions$>$} $\to$ {\it $<$Action$>$}.
In {\it $<$Conditions$>$}, 
we use a triplet (ordered triple) $(\XX,\YY,\Co)$, 
which means that the number of robots satisfying the following three conditions among all observed robots: 
$\XX$ (resp. $\YY$) indicates the range of x-coordinate (resp. y-coordinate), 
and $\Co$ indicates the color of the light: $\Co \in \CC \cup \{ \CS \}$, where $\CS$ means any color 
(\ie do not care).
{\it $<$Action$>$} is presented as a tuple $(a^d,\CC)$:
the former one $a^d$ represents the destination point to move and
the latter one $\CC$ represents the color of the light to change.
Remind that each robot can move to its adjacent (four) point, 
thus we can present the destination point using $a^d$, where $a \in \{x,y\}$ and $d \in \{+,-\}$, 
which means the orientation and the direction respectively based on its local coordination system.
Both $a^d$ and $\CC$ can have $\bot$, which means null movement (does not move) and 
keep color (does not change color) respectively.
We also use a macro $\myC$ which indicates its own light.
As a common rule that applies to all rules,
a robot never moves if the destination point is already occupied by another robot, 
even if the conditions are satisfied.

Here we give an example of the algorithm to help to understand.

\begin{algorithm}
    \caption{An example of Algorithm}
    \label{exalgo}
    \begin{algorithmic}
        \STATE {\bf Rule $x$:} $\myC = \CA \land (3 \geq x \geq 1, y = 2, \CA) = 3 \land ( x = 0, y \geq 1, \CS) = 1 \to (x^+, \CB)$
    \end{algorithmic}
\end{algorithm}

Algorithm \ref{exalgo} shows the $x$-th rule:
robot $r_i$ changes the color of its light to $\CB$ (in the end of its {\sl Compute} phase) 
and moves to the positive direction of X-axis (in its {\sl Move} phase)
if all the following conditions are satisfied;
(1) the color of its light is $\CA$, 
(2) the number of the observed robots,
whose x-coordinate is between 1 and 3 (based on $r_i$'s local coordination),
y-coordinate is exactly 2 (based on $r_i$'s local coordination), and color of the light is $\CB$,
is equal to 3, 
and
(3) there exists a robot on the positive direction of $r_i$'s current Y-axis.

It is worthwhile to mention that different algorithms are required where the number of robots $n$ is 2, 3, and 4 or more.
We present the algorithm only for $n \geq 4$, which is the most general case. 
Note that the number of robots can be counted in the coordinate system generation phase, 
thus we can determine the algorithm to execute based on $n$.

\subsection{Phase 1: The Line Formation Phase}
Algorithm \ref{algo_phase1} shows the first phase of 
the proposed algorithm, called \emph{the Line Formation Phase}, 
which consists of 15 rules, from Rule 1 to Rule 15.

\begin{algorithm}[tbhp]
    \caption{The Line Formation Phase}
    \begin{algorithmic}
\STATE {\it /* From an initial configuration to configuration $C_A$ */}
        \STATE {\bf Rule 1:} $\myC = \CA \ \land \ (x \leq -1, y \leq -1, \CS) = 0 \ \land \ (x \geq 1, y \leq -1, \CS) = 0 
        \land \ (x = 0, y \leq -1, \CS) = 0 \ \land \ (x \geq 1, y = 0, \CS) = 0 
        \land \ (x \leq -1, y = 0, \CA) = 1 \ \land \ (-\infty \leq x \leq \infty,-\infty \leq y \leq \infty, \CB) = 0
        \to (y^-, \bot)$
       \STATE {\bf Rule 2:} $\myC = \CA \land (x \leq -1, y \leq -1, \CS) = 0 \land (x \geq 1, y \leq -1, \CS) = 0 \land (x = 0, y \leq -1, \CS) = 0 \land
                                  (x \geq 1, y = 0, \CS) = 0 \land (x \leq -1, y = 0, \CS) = 0 \land ((x = 0, y \geq 1, \CA) \geq 1 \lor (x \geq 1, y \geq 1, \CA) \geq 1) \land
                                  (-\infty \leq x \leq \infty,-\infty \leq y \leq \infty, \CB) = 0 \to (x^+, \bot)$
       \STATE {\bf Rule 3:} $\myC = \CA \land (x \leq -1, y \geq 1, \CA) \geq 1 \land (\lnot(x \leq -1), \lnot(y \geq 1), \CS) = 0
                                   \to (\bot, \CB)$
\STATE {\it /* From configuration $C_A$ to configuration $C_B$ */}
       \STATE {\bf Rule 4:} $\myC = \CA \land (x \leq 1, y \leq -1, \CB) = 1 \land 
                                  (\lnot(x \geq 1), \lnot(y \leq -1), \CB) =0 \to (x^+, \bot)$
       \STATE {\bf Rule 5:} $\myC = \CA \land (x = 1, y \leq -1, \CB) = 1 \land
                                  (\lnot(x = 1), \lnot(y \leq -1), \CB) = 0 \land (-\infty \leq x \leq \infty, y \leq -1, \CA) = 0 \to (x^+,\bot)$
       \STATE {\bf Rule 6:} $\myC = \CA \land (x = 0, y \leq -1, \CB) = 1 \land
                                  (-\infty \leq x \leq \infty, \lnot(y \leq -1), \CB) = 0 \land (-\infty \leq x \leq \infty, y \leq -1, \CA) = 0 \land
                                  (x \geq 0, y \geq 0, \CA) = 0 \to (\bot,\CB)$
\STATE {\it /* From configuration $C_B$ to configuration $C_C$ */}
      \STATE {\bf Rule 7:} $\myC = \CB \land (x \geq 1, -\infty \leq y \leq \infty, \CS) = 0 \land (x = 0, y \geq 1, \CS) = 0 \land
                                 (x \leq -1, y \geq 0, \CA) \geq 1 \to (y^+,\bot)$
      \STATE {\bf Rule 8:} $\myC = \CB \land (x \geq 1, -\infty \leq y \leq \infty, \CS) = 0 \land (x = 0, y \leq -1, \CS) = 0 \land
                                 (x \leq -1, y \leq 0, \CA) \geq 1 \to (y^-,\bot)$
      \STATE {\bf Rule 9:} $\myC = \CA \land 1 \leq (x \geq 2, -\infty \leq y \leq \infty, \CB) \leq 2 \land
                                (x \leq 1, -\infty \leq y \leq \infty, \CB) = 0 \to (x^+,\bot)$
      \STATE {\bf Rule 10:} $\myC = \CA \land (x = 1, y \geq 1, \CB) = 1 \land
                                (x = 1, y \leq -1, \CB) = 1 \land (x \leq 0, -\infty \leq y \leq \infty, \CB) = 0
                                 \to (x^+,\bot)$
      \STATE {\bf Rule 11:} $\myC = \CA \land (x = 1, y \geq 1, \CB) = 1 \land
                                (x = 1, y \leq -1, \CB) = 1 \land (x \leq 0, -\infty \leq y \leq \infty, \CB) = 0 \land
                                 (x = 1, y = 0, \CA) = 1 \land (x \leq 0, y \geq 1, \CS) = 0
                                 \to (y^+,\bot)$
      \STATE {\bf Rule 12:} $\myC = \CA \land (x = 1, y \geq 1, \CB) = 1 \land
                                (x = 1, y \leq -1, \CB) = 1 \land (x \leq 0, -\infty \leq y \leq \infty, \CB) = 0\land
                                 (x = 1, y = 0, \CA) = 1 \land (x \leq 0, y \leq -1, \CS) = 0
                                 \to (y^-,\bot)$
      \STATE {\bf Rule 13:} $\myC = \CA \land (x = 1, y \geq 1, \CB) = 1 \land
                                (x = 1, y \leq -1, \CB) = 1 \land (x \leq 0, -\infty \leq y \leq \infty, \CB) = 0 \land
                                 (x = 1, y = 0, \CA) = 1 \land (x \leq 0, y \leq -1, \CS) = 0 \land
                                 (x \leq 0, y \geq 1, \CS) = 0 \to (y^+,\bot)$
\STATE {\it /* From configuration $C_C$ to configuration $C_D$ */}

      \STATE {\bf Rule 14:}$(x = 0, y \geq 2, \CS) = 1 \land (x \geq 1, -\infty \leq y \leq \infty, \CS) = 0 \land
                                  (x \leq -1, -\infty \leq y \leq \infty, \CS) = 0 \to (y^+,\bot)$
      \STATE {\bf Rule 15:} $\myC = \CA \land (x = 0, y = 1, \CB) = 1 \land (x \geq 1, -\infty \leq y \leq \infty, \CS) = 0 \land
                                  (x \leq -1, -\infty \leq y \leq \infty, \CS) = 0 \to (\bot,\CB)$
    \end{algorithmic}
    \label{algo_phase1}
\end{algorithm}

In this phase, we consider four different configurations, called configurations from $C_A$ to $C_D$ respectively, as the following.

\noindent
{\bf Configuration $C_A$}:  
Only one robot with color $\CB$ is located at a point with the smallest y-coordinate and the largest x-coordinate and all other robots have lights with color $\CA$.

\noindent
{\bf Configuration $C_B$}: 
Two robots with color $\CB$ are on the same Y-axis and can observe each other (\ie there is no other robot between the two robots). Moreover, all other robots than the two robots with color $\CB$ have lights with color $\CA$.
The two robots with color $\CB$ are located on the same or smaller y-coordinate than any other robots with color $\CA$.

\noindent
{\bf Configuration $C_C$}: 
All robots are located on the same Y-axis, and the robot with the smallest y-coordinate and the robot with the largest y-coordinate have lights with color $\CB$, while all other robots have lights with color $\CA$.

\noindent
{\bf Configuration $C_D$}: 
All robots are located on the same Y-axis, have the same lights with color $\CB$, and are lined up without gaps.

\begin{figure}[tbhp]
	\begin{center}
		\subfloat[$C_A$]{\includegraphics[scale=0.35]{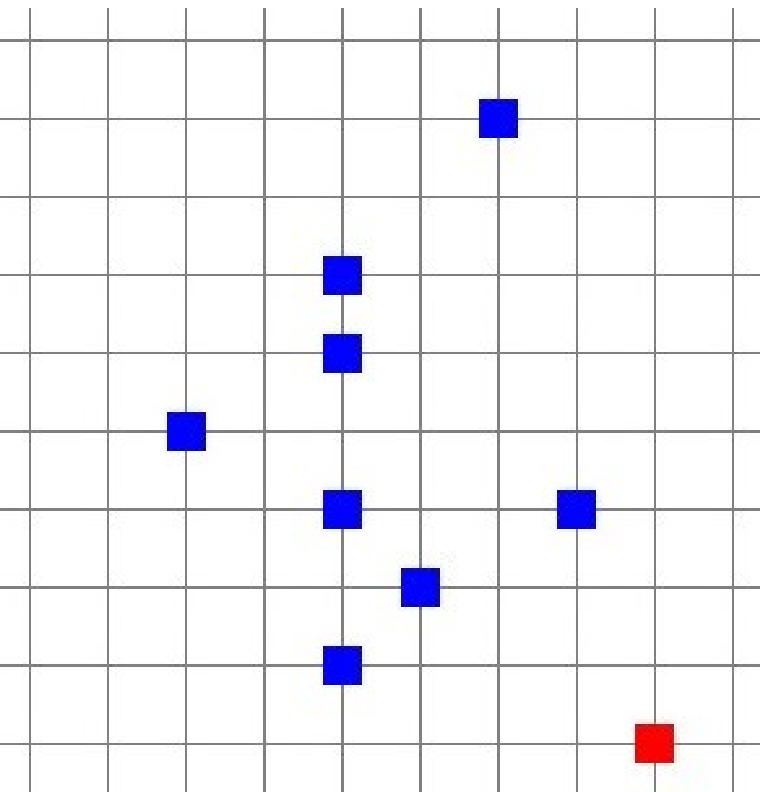}}
        \hspace{5pt}
		\subfloat[$C_B$]{\includegraphics[scale=0.35]{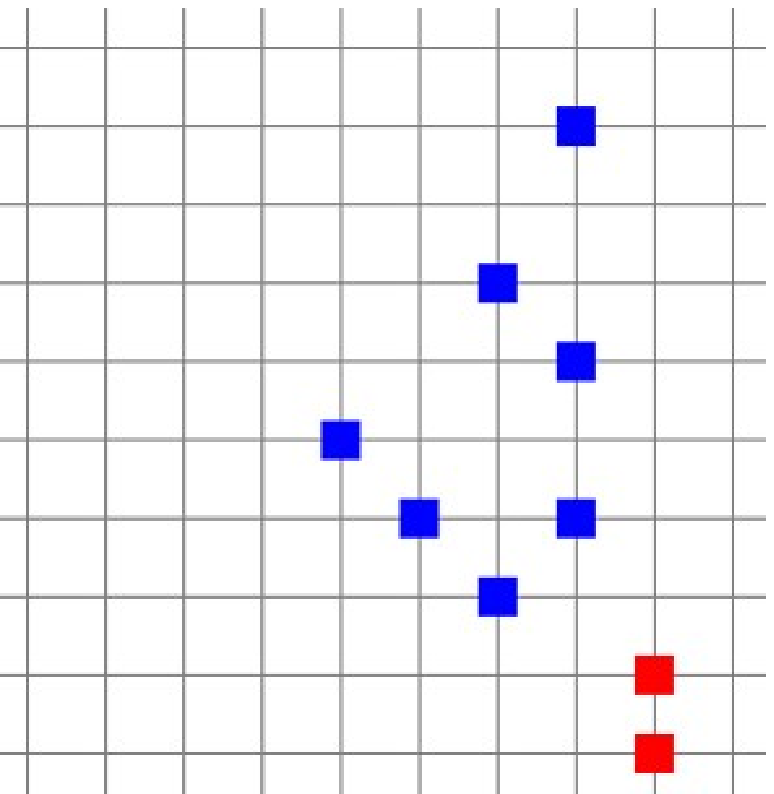}}
        \hspace{5pt}
		\subfloat[$C_C$]{\includegraphics[scale=0.35]{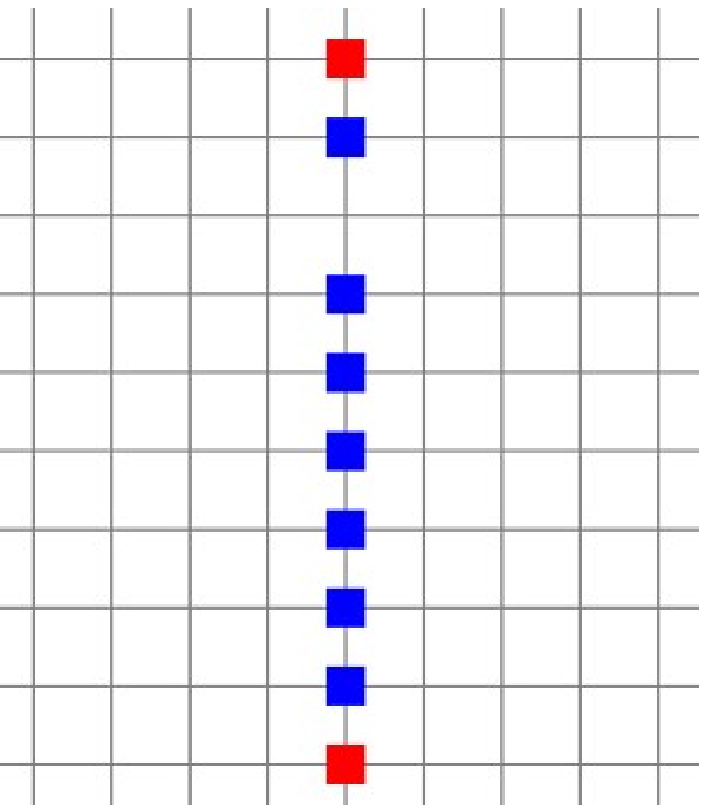}}
        \hspace{5pt}
		\subfloat[$C_D$]{\includegraphics[scale=0.35]{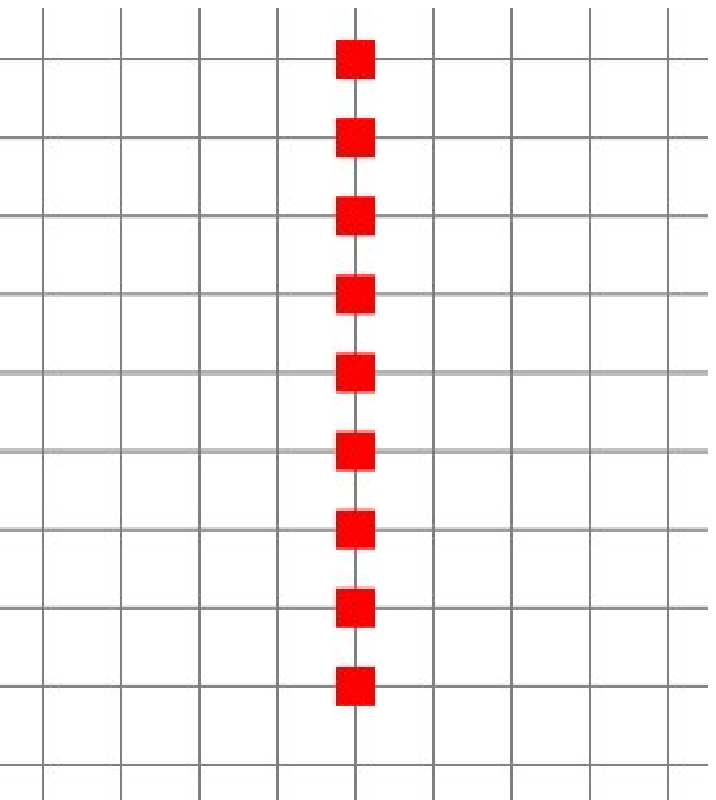}}
	\end{center}
    \caption{Examples of configurations from $C_A$ to $C_D$}
    \label{fig:confCAtoCD}
\end{figure}

Figure \ref{fig:confCAtoCD}(a) to Figure \ref{fig:confCAtoCD}(d) illustrate the examples of these four configurations 
in the {\em Line Formation Phase} respectively. 
First of all, the proposed algorithm determines the reference point to form a straight line from any initial configuration
so that all the robots have light with color $\CA$. 
The proposed algorithm uses the robot which has the largest x-coordinate and the smallest y-coordinate, 
\ie the rightmost and undermost robot, as the reference point 
to determine where the robots are lined up.

\begin{figure}[tbhp]
	\begin{center}
		\subfloat[Rule 1]{\includegraphics[scale=0.35]{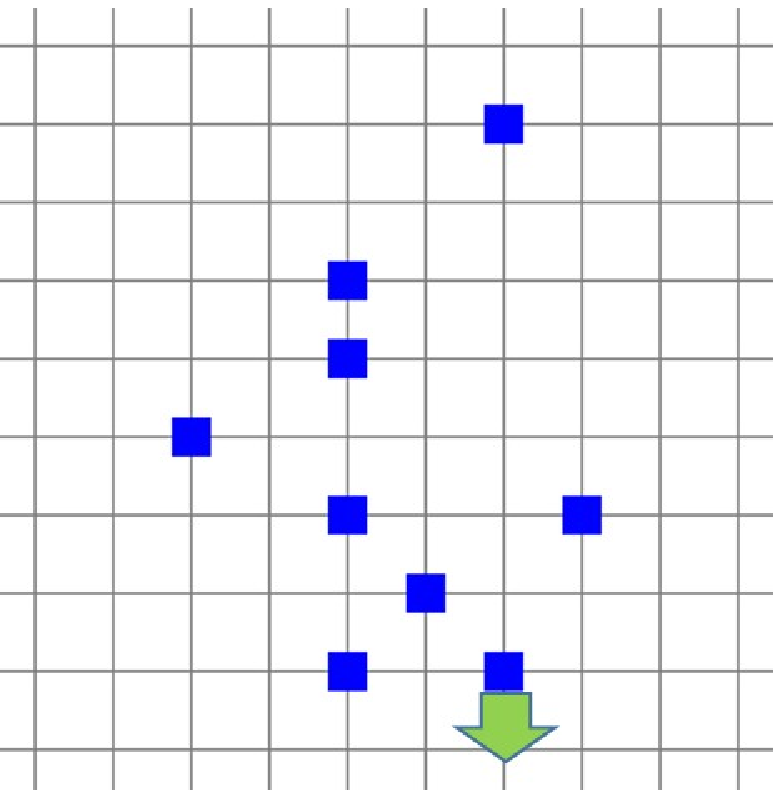}}
        \hspace{25pt}
		\subfloat[Rule 2]{\includegraphics[scale=0.35]{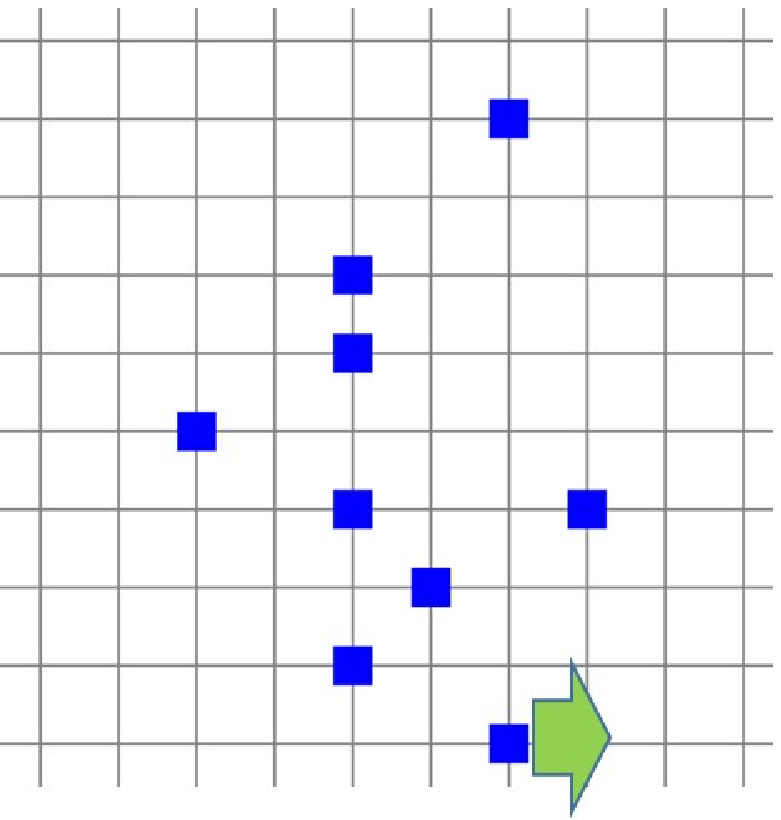}}
        \hspace{25pt}
		\subfloat[Rule 3]{\includegraphics[scale=0.35]{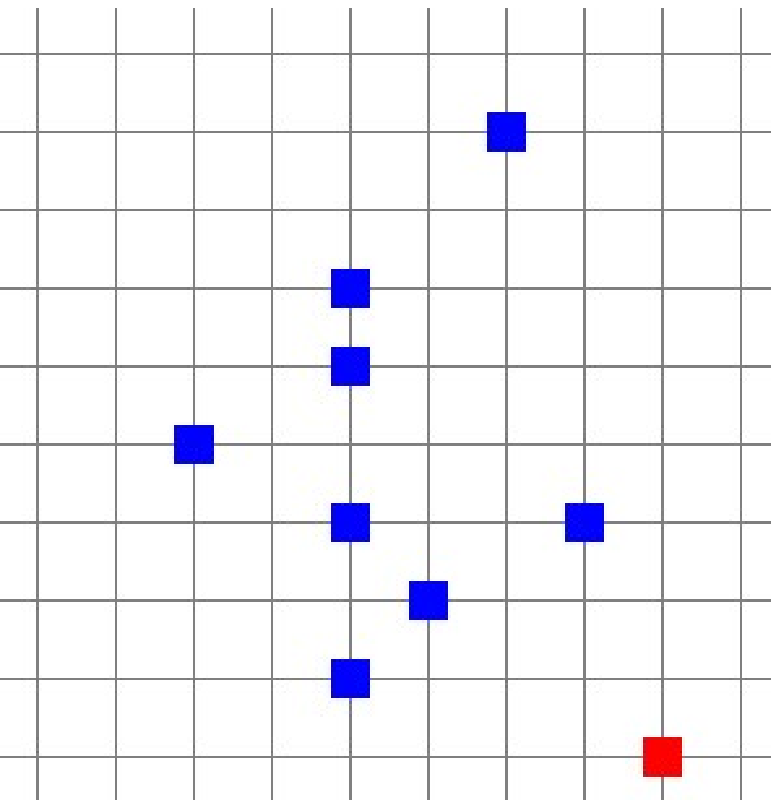}}
	\end{center}
    \caption{Examples of movements by Rules 1 to 3}
    \label{fig:Rule123}
\end{figure}

The proposed algorithm uses 3 rules to achieve configuration $C_A$ 
from any initial configuration 
(refer to Rules 1 to 3 in Algorithm \ref{algo_phase1}).
In any initial configuration, 
there exists a robot that cannot observe any other robots in the third and forth quadrants based on its local coordinate
system by assuming its current position as the origin, \ie there is no robot located on the point with the smaller y-coordinate
than the robot's.
If the robot cannot observe any other robots in positive direction of the same X-axis neither,
the robot becomes the reference robot (denoted by $\alpha$). And robot $\alpha$ moves to the appropriate reference point.
If robot $\alpha$ can observe another robot in its negative direction of the same X-axis, it first moves to the negative
direction along the current Y-axis (Figure \ref{fig:Rule123}(a)).
If robot $\alpha$ cannot observe any other robots on the negative direction of its current X-axis, robot $\alpha$ moves to the positive direction along the X-axis (Figure \ref{fig:Rule123}(b)).
If robot $\alpha$ cannot observe any other robots both in the negative direction of its current X-axis 
and in its first quadrant, 
it changes its color of the light to $\CB$, because the robot reaches the reference point (Figure \ref{fig:Rule123}(c)).
This implies that robot $\alpha$ changes its color from $\CA$ to $\CB$ 
only when it observes all other robots with color $\CB$ in its second quadrant. 
When robot $\alpha$ changes its color to $\CB$, configuration $C_A$ is achieved.


\begin{figure}[tbhp]
	\begin{center}
		\subfloat[Rule 4]{\includegraphics[scale=0.35]{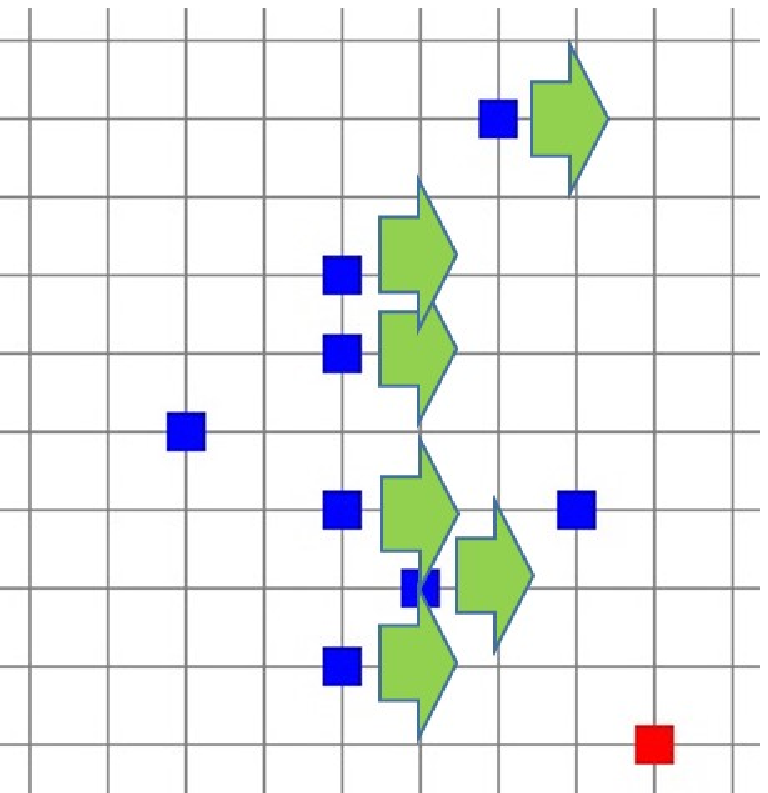}}
        \hspace{25pt}
		\subfloat[Rule 5]{\includegraphics[scale=0.35]{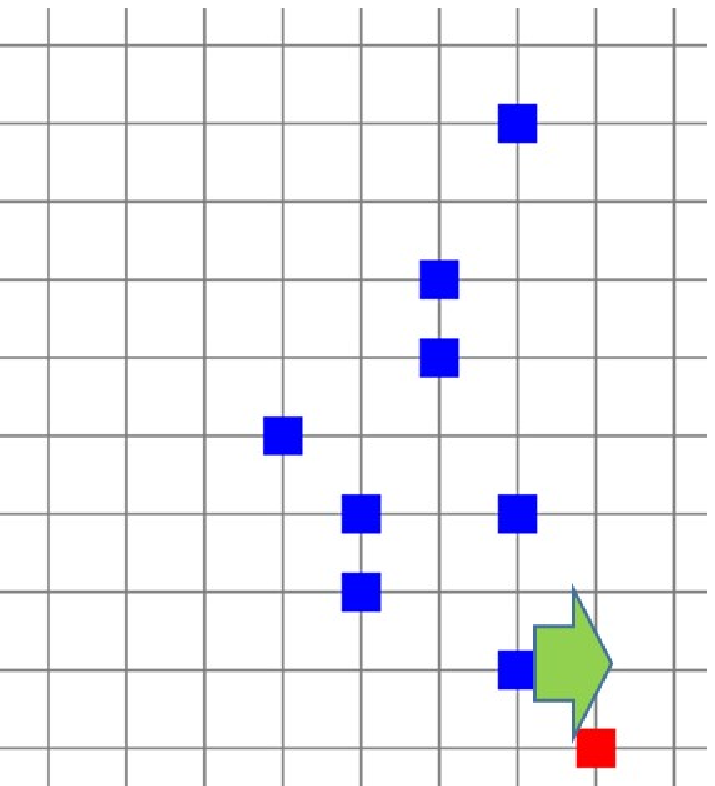}}
        \hspace{25pt}
		\subfloat[Rule 6]{\includegraphics[scale=0.35]{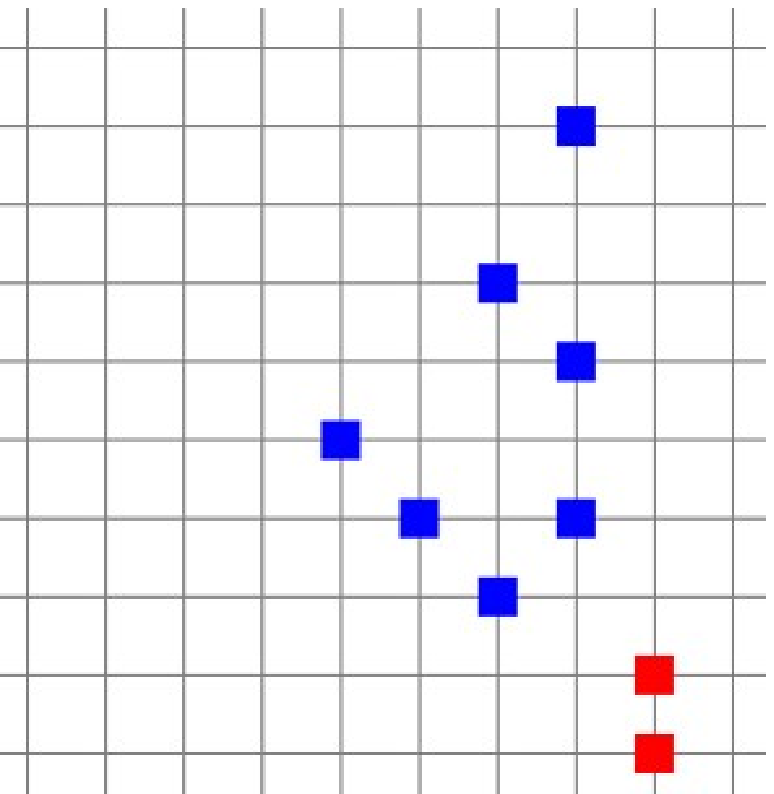}}
	\end{center}
    \caption{Examples of movements by Rules 4 to 6}
    \label{fig:Rule456}
\end{figure}

The proposed algorithm uses 3 rules to achieve configuration $C_B$ 
from configuration $C_A$ 
(refer to Rules 4 to 6 in Algorithm \ref{algo_phase1}).
Let $y_0$ be the Y-axis where robot $\alpha$ exists in configuration $C_A$,
and each $y_i$ be the Y-axis $i$ away from $y_0$ on the positive or negative direction of X-axis.
For example, $y_1$ represents the Y-axis whose x-coordinate is 1 larger than $y_0$, \ie the right neighbor of $y_0$,
$y_{-1}$ represents the Y-axis whose x-coordinate is 1 smaller than $y_0$, \ie the left neighbor of $y_0$.
In configuration $C_A$, robots with color $\CA$ can observe only one robot with color $\CB$ in its forth quadrant, 
or no robot with color $\CB$ (because they are not transparent).
The robot with color $\CA$, which observes a robot with color $\CB$ in its forth quadrant,
moves to the positive direction of its current X-axis toward $y_{-1}$ 
(Figure \ref{fig:Rule456}(a)).
Note that each robot moves only to the positive direction of X-axis \ie right, 
and a robot does not move to its destination point if another robot exists there 
to avoid a collision.
Therefore, each robot with color $\CB$ stops when it reaches $y_{-1}$ or 
there exists another robot on its adjacent point on the positive direction of X-axis
(if it can observe robot $\alpha$).

Now we consider a robot with color $\CA$ which reaches $y_{-1}$ and 
cannot observe any other robot on its negative side of Y-axis.
%
We call this robot $\beta$ and only robot $\beta$ moves to $y_0$ 
whereas all other robots with color $\CA$ stops when they reaches $y_{-1}$.
After that, robot $\beta$ changes its light color to $\CB$ (Figure \ref{fig:Rule456}(c)).
When robot $\beta$ changes the light color to $\CB$, configuration $C_B$ is achieved.

\begin{figure}[tbhp]
	\begin{center}
		\subfloat[Rules 7 and 8]{\includegraphics[scale=0.35]{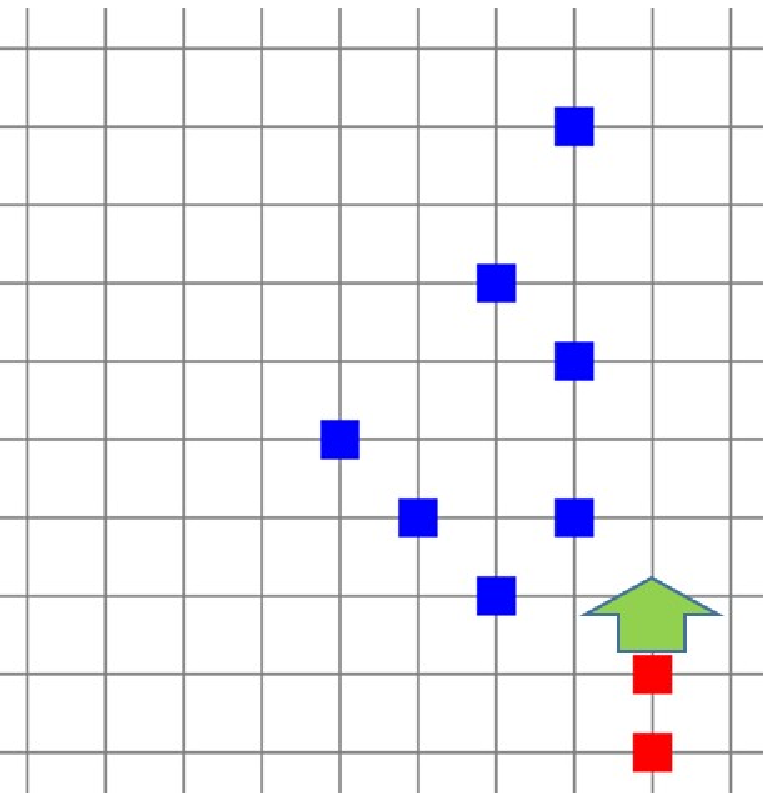}}
        \hspace{5pt}
		\subfloat[Rule 9]{\includegraphics[scale=0.35]{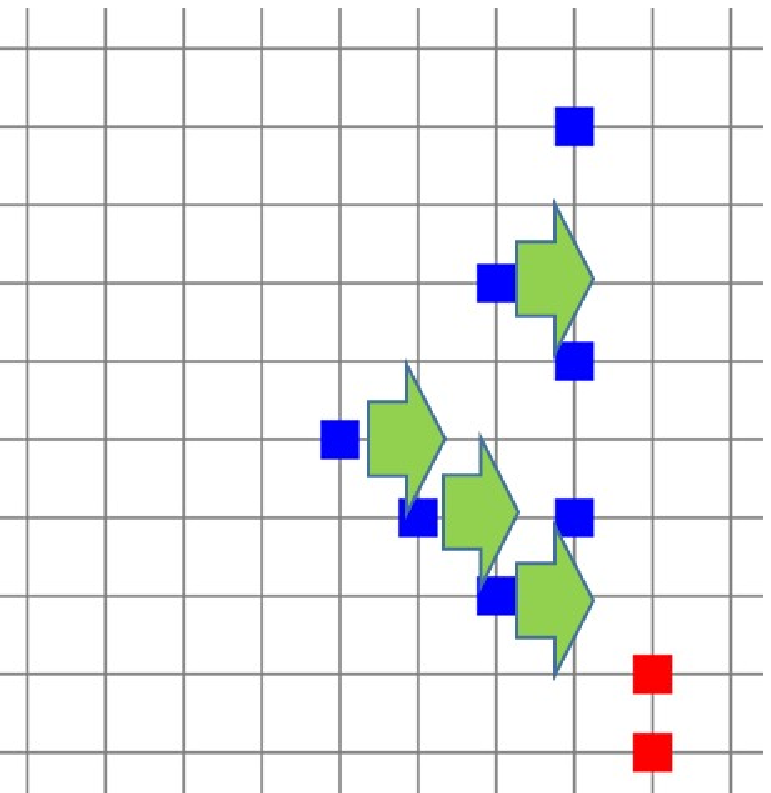}}
        \hspace{5pt}
		\subfloat[Rule 10]{\includegraphics[scale=0.35]{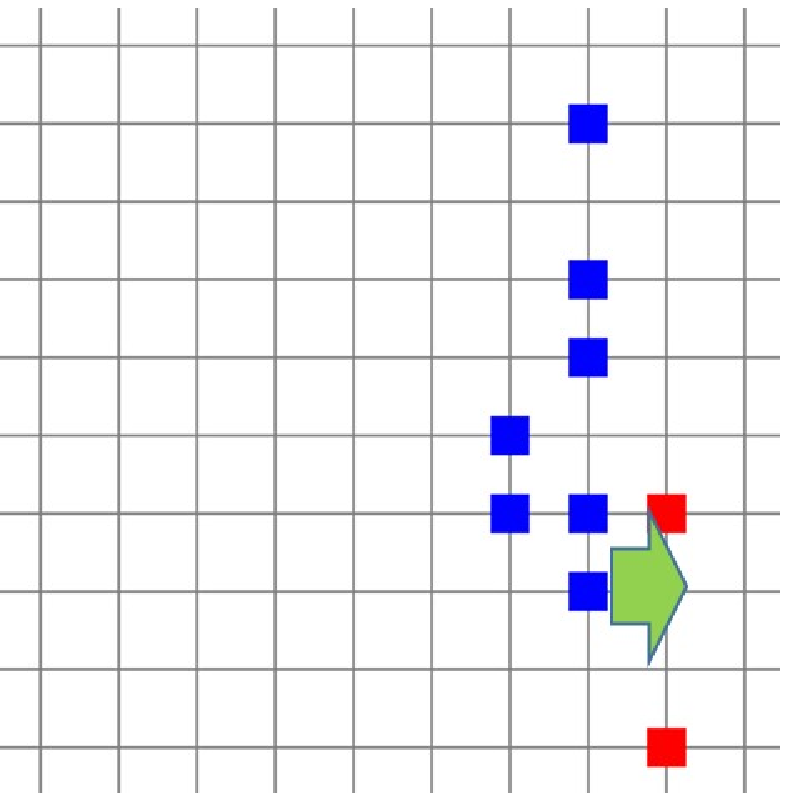}}
        \hspace{5pt}
		\subfloat[Rules 11 to 13]{\includegraphics[scale=0.35]{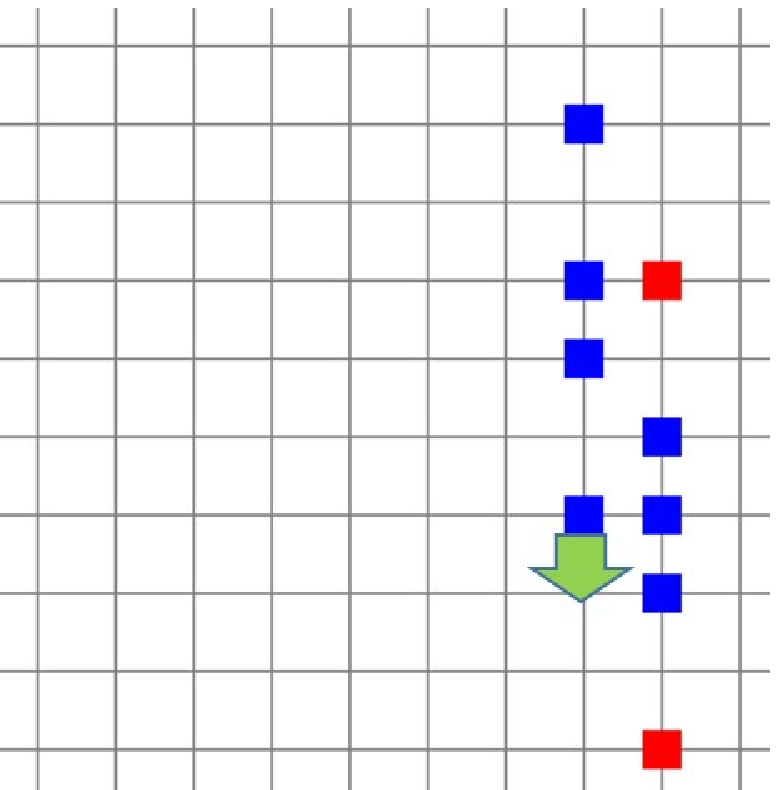}}
	\end{center}
    \caption{Examples of movements by Rules 7 to 13}
    \label{fig:Rule7to13}
\end{figure}

The proposed algorithm uses 7 rules to achieve configuration $C_C$ 
from configuration $C_B$ 
(refer to Rules 7 to 13 in Algorithm \ref{algo_phase1}).
In configuration $C_B$, two robots with color $\CB$ (robots $\alpha$ and $\beta$) exist on the same Y-axis and they can observe each other.
Robot $\alpha$ has a smaller y-coordinate than all the other robots, 
and robot $\beta$ has a equal or smaller y-coordinate than all the other robots except robot $\alpha$.
Note that, robots $\alpha$ and $\beta$ have larger x-coordinate than all other robots.
Our strategy in this phase is to locate all other robots than $\alpha$ and $\beta$ at the points 
between two robots $\alpha$ and $\beta$.
To use this strategy, robots $\alpha$ and $\beta$ should move away from each other if necessary.
Notice that, in Configuration $B$, only two robots $\alpha$ and $\beta$ have a light with color $\CB$ 
and all other robots are located on the negative side of X-axis in local coordinates of robots $\alpha$ and $\beta$ 
(Figure \ref{fig:Rule7to13}(a)).
If a robot with color $\CA$ observes one or two robots with color $\CB$ on its positive side of X-axis, 
it moves along its current X-axis toward the $y_{-1}$ (Figure \ref{fig:Rule7to13}(b)).
If a robot on $y_{-1}$ observes one robot with color $\CB$ in the first quadrant 
and another robot with color $\CB$ in forth quadrant, 
it moves to its neighboring point on $y_0$ (Figure \ref{fig:Rule7to13}(c)) if the point is empty.
%
However, if there is another robot on its neighboring point in $y_0$, 
it moves along its Y-axis according to the following rules:
if there is no robot in the positive (resp. negative) direction of its current Y-axis and 
the second (resp. third) quadrant based on its local coordinate system,
the robot moves to the positive (resp. negative) direction of its Y-axis
(Figure \ref{fig:Rule7to13}(d)).
Note that if it is possible to move to both directions, the robot moves to the positive direction of Y-axis 
due to the priority of the algorithm.
When all robots are located on $y_0$, the configuration satisfies configuration $C_C$.

\begin{figure}[tbhp]
 \begin{minipage}{0.65\hsize}
	\begin{center}
		\subfloat[Rule 14]{\includegraphics[scale=0.34]{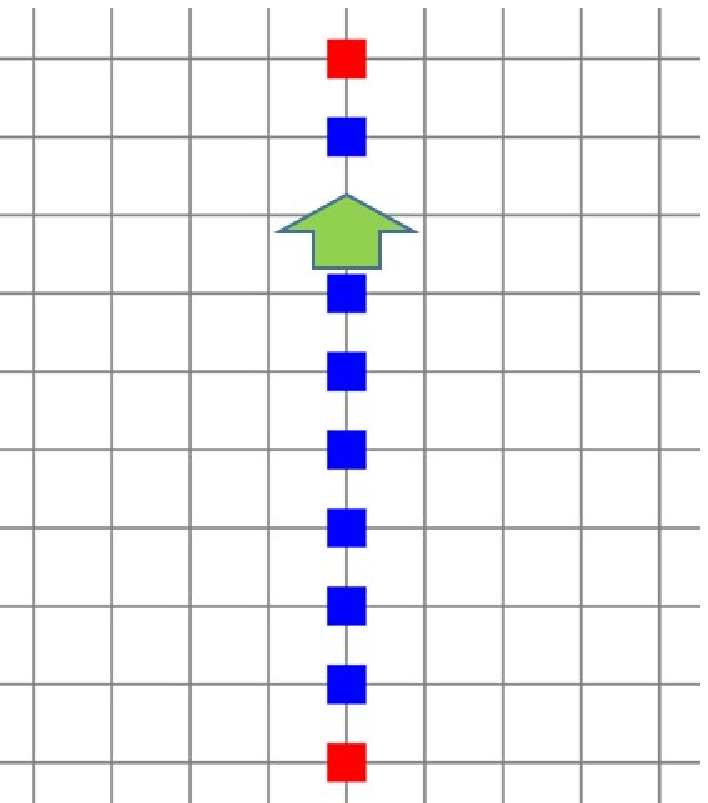}}
        \hspace{20pt}
		\subfloat[Rule 15]{\includegraphics[scale=0.34]{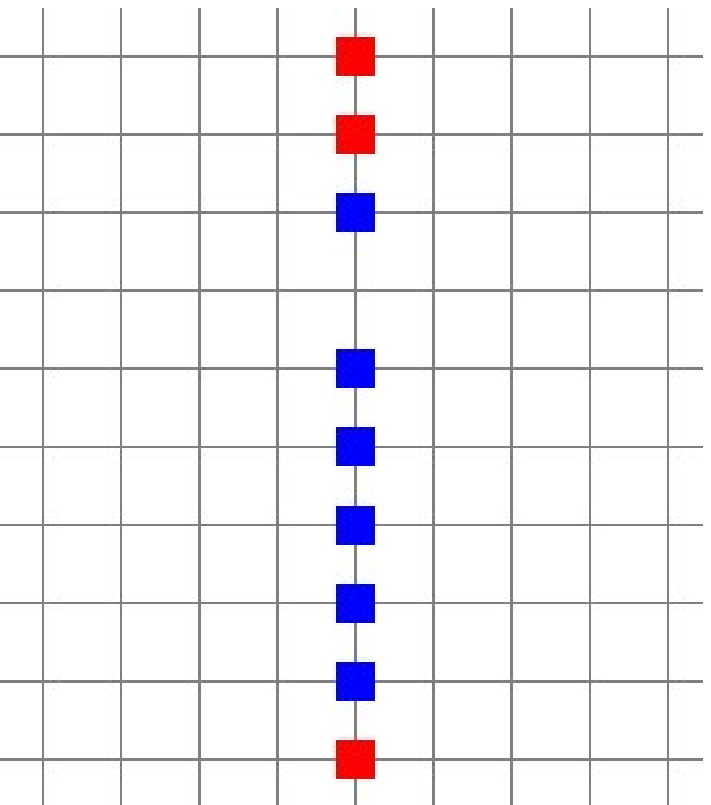}}
	\end{center}
    \caption{Examples of movement by Rules 14 and 15}
    \label{fig:Rule1415}
 \end{minipage}
 \begin{minipage}{0.29\hsize}
	\begin{center}
		\includegraphics[scale=0.37]{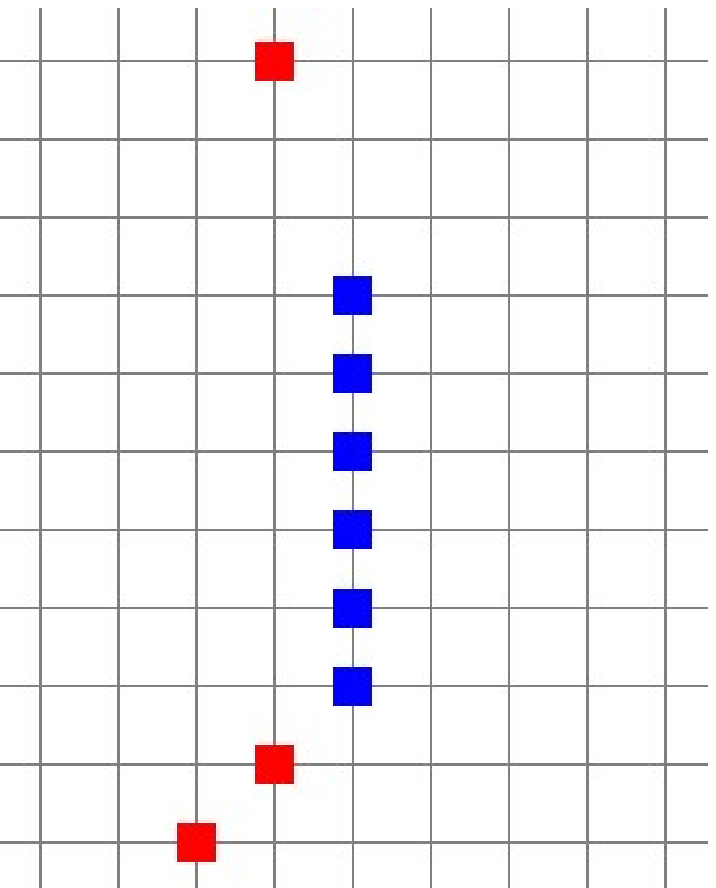}
	\end{center}
  \caption{$C_E$ when $n=9$}
  \label{fig_confCE}
 \end{minipage}
\end{figure}

The proposed algorithm uses two rules to achieve configuration $C_D$ 
from configuration $C_C$ 
(refer to Rule 14 and Rule 15 in Algorithm \ref{algo_phase1}).
In configuration $C_C$, all robots are located on the same Y-axis, 
and only two robots with the smallest and largest y-coordinate respectively (robots $\alpha$ and $\beta$) 
have light with color $\CB$, 
while all the other robots have light with color $\CA$.
Note that, in this case, all robots observe no robot in all quadrants and on its current X-axis.
If a robot observes an other robot in the positive direction of its Y-axis 
but no robot on its neighboring point in the positive direction of its Y-axis, 
the robot moves to the positive direction of its Y-axis to close the gap (Figure \ref{fig:Rule1415}(a)).
If a robot with color $\CA$ observes in the adjacent point in the positive direction of the Y-axis, 
it changes its color into $\CB$ (Figure \ref{fig:Rule1415}(b)).
If all robots are lined up without gaps, configuration $C_D$ is achieved, 
and the first phase, {\em The Line Formation Phase}, is completed.

\subsection{Phase 2: The Coordinate System Generation Phase}
Algorithm \ref{algo_phase2} presents the second phase of 
the proposed algorithm, called \emph{the Coordinate System Generation Phase}, 
which consists of 8 rules, from Rule 16 to Rule 23.

\begin{algorithm}[tbhp]
    \caption{Coordinate System Generation Phase}
    \begin{algorithmic}
\STATE {\it /* From configuration $C_D$ to configuration $C_E$ */}
       \STATE {\bf Rule 16:} $\myC = \CB \land (x = 0, y = 1, \CB) = 1 \land (\lnot(x = 0, y = 1), \CS) = 0
                                  \to (x^-,\bot)$
       \STATE {\bf Rule 17:} $\myC = \CB \land (x = -1, y \leq -1, \CB) = 1 \land (-\infty \leq x \leq \infty, y = 0, \CS) = 0
                                  \to (\bot,\CA)$
       \STATE {\bf Rule 18:} $\myC = \CB \land (x = 1, y \geq 1, \CA) \geq 3 \land (\lnot (x = 1, y \geq 1), \CS) = 0
                                  \to (x^-,\bot)$
       \STATE {\bf Rule 19:} $\myC = \CA \land (x = -2, y = -1, \CB) = 1 \land (x = 0, y = 1, \CA) = 1 \land  (\lnot((x = -2, y = -1) \land (x = 0, y = 1)), \CS) = 0
                                  \to (x^-,\bot)$
       \STATE {\bf Rule 20:} $\myC = \CA \land (x = -1, y = -1, \CB) = 1 \land (x = 1, y \geq 1, \CA) \geq 2 \land  (\lnot((x = -1, y = -1) \land (x = 1, y \geq 1)), \CS) = 0
                                  \to (\bot,\CB)$
       \STATE {\bf Rule 21:} $\myC = \CA \land (x = -1, y \leq -1, \CB) = 1 \land (x = -2, y \leq -1, \CB) = 1 \land (x = 0, y = -1, \CA) = 1 \land
                                  (\lnot((x = -1, y \leq -1) \land (x = -2, y \leq -1) \land (x = 0, y = -1)), \CS) = 0 \to (x^-,\bot)$\\
       \STATE {\bf Rule 22:} $\myC = \CA \land (x = 0, y \leq -1, \CB) = 1 \land (x = -1, y \leq -1, \CB) = 1 \land (x = 0, y = -m + 1, \CB) = 0 \land (\lnot(x = 0, y \leq -1) \land (x = -1, y \leq -1), \CB) = 0 \to (y^+,\bot)$
       \STATE {\bf Rule 23:} $\myC = \CA \land (x = 0, y \leq -1, \CB) = 1 \land (x = -1, y \leq -1, \CB) = 1 \land (x = 0, y = -m + 1, \CB) = 1 \land (\lnot(x = 0, y \leq -1) \land (x = -1, y \leq -1), \CB) = 0 \to (\bot,\CB)$

    \end{algorithmic}
    \label{algo_phase2}
\end{algorithm}

Here we define configuration $C_E$ which is the goal configuration of this phase as the following:

\noindent
{\bf Configuration $C_E$}: 
Let a robot with color $\CB$ with the smallest y-coordinate be the origin, 
  there exists two robots with color $\CB$ at (1,1) and (1,$m$-1) respectively, 
  where $m$ is the smallest prime number greater than or equal to $n$. 
  All other robots have light with color $\CA$ and are lined up with no gaps on the line, $y=2$, from (2,2) to (2,$n$-2)
  (Figure \ref{fig_confCE}).

In this phase, only some specific robots can operate (moves or changes the color of light) in every configuration, 
thus the goal configuration is achieved by the sequential operations of the robots.
The proposed algorithm uses 8 rules to achieve the second phase
(refer to Algorithm \ref{algo_phase2}).

\begin{figure}[tbhp]
	\begin{center}
		\subfloat[Rule 16]{\includegraphics[scale=0.35]{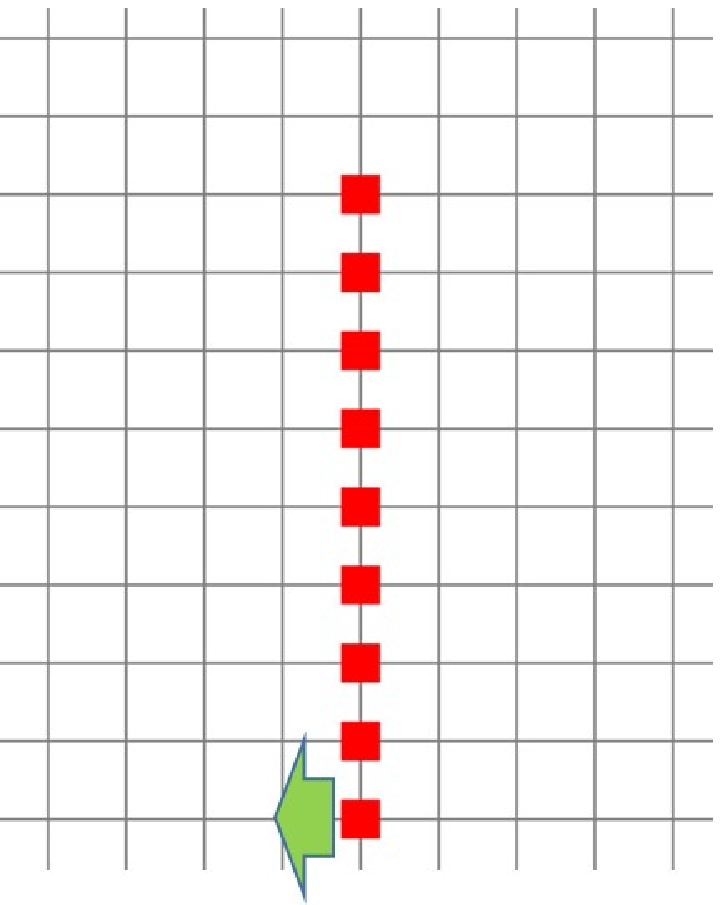}}
        \hspace{10pt}
		\subfloat[Rule 17]{\includegraphics[scale=0.35]{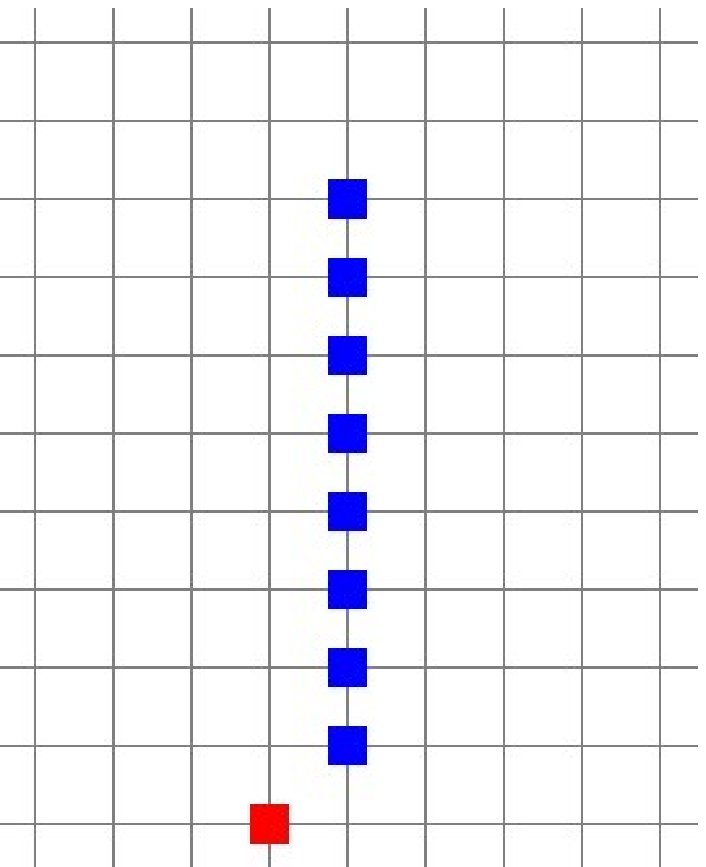}}
        \hspace{10pt}
		\subfloat[Rule 18]{\includegraphics[scale=0.35]{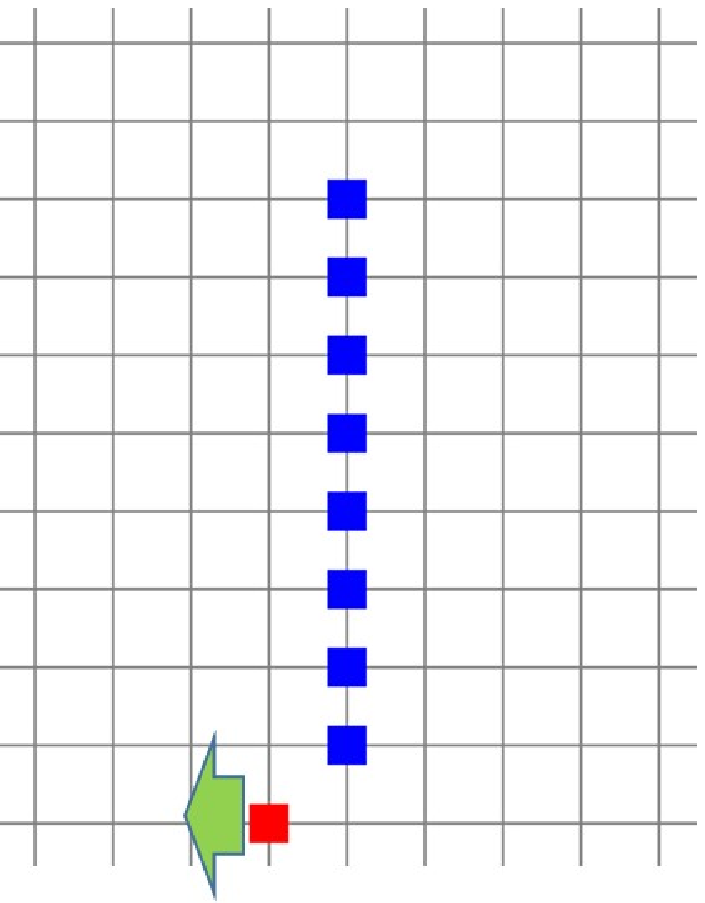}}
        \hspace{10pt}
		\subfloat[Rule 19]{\includegraphics[scale=0.35]{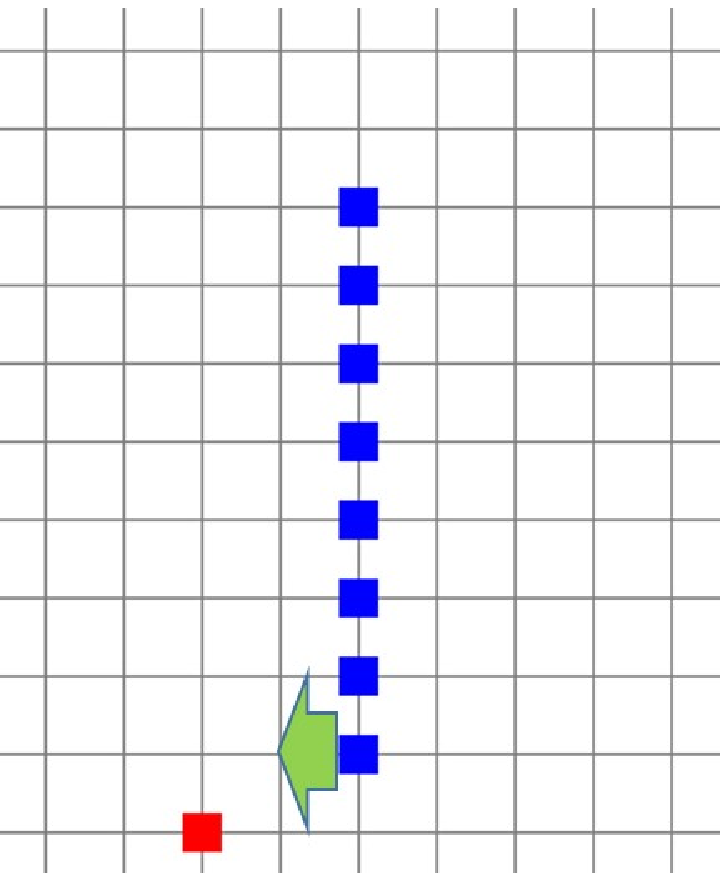}}
	\end{center}
    \caption{Examples of movements by Rules 16 to 19}
    \label{fig:Rule16to19}
\end{figure}

\begin{figure}[tbhp]
	\begin{center}
		\subfloat[Rule 20]{\includegraphics[scale=0.35]{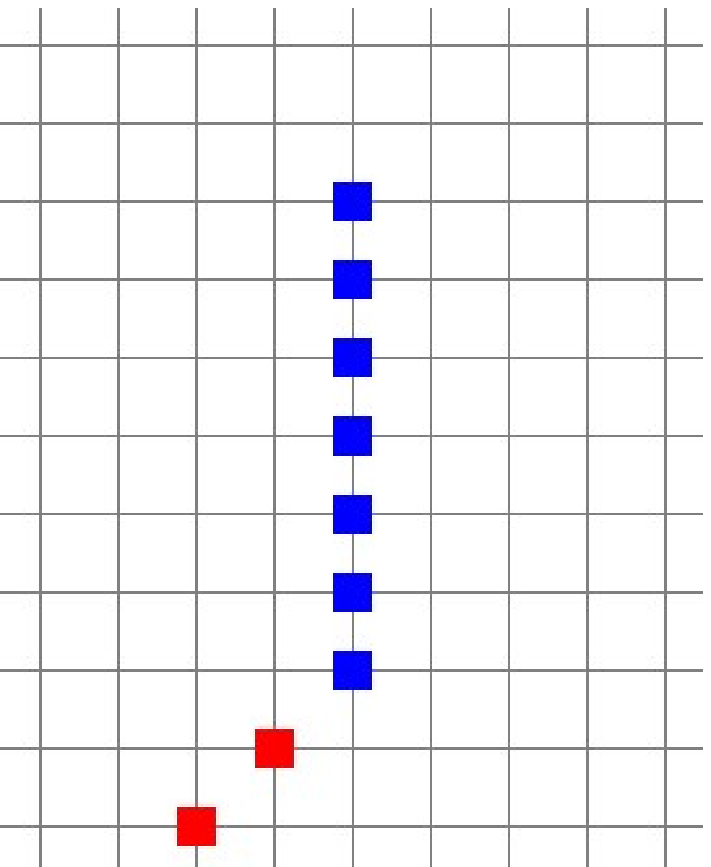}}
        \hspace{10pt}
		\subfloat[Rule 21]{\includegraphics[scale=0.35]{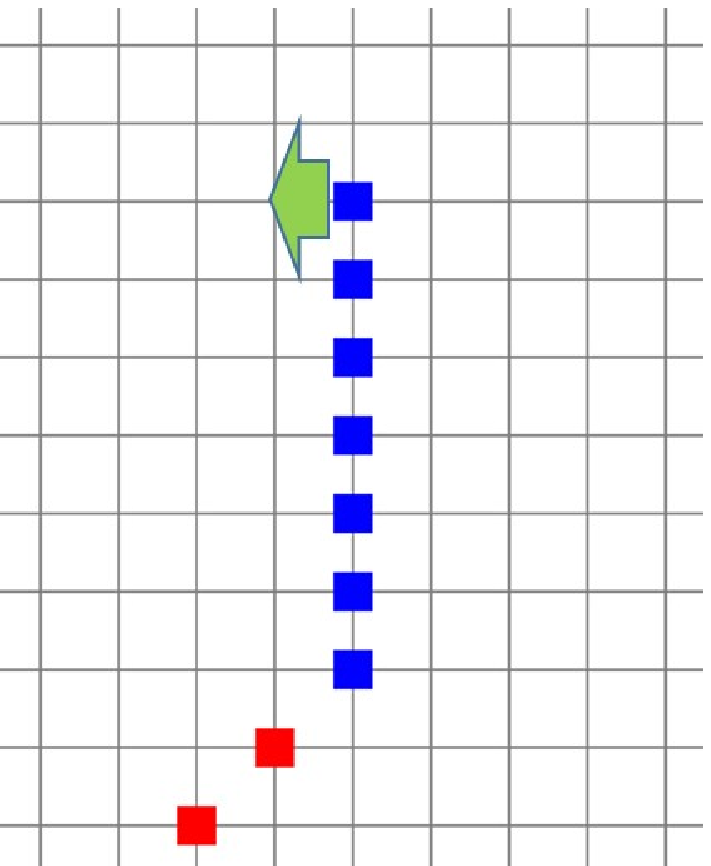}}
        \hspace{10pt}
		\subfloat[Rule 22]{\includegraphics[scale=0.35]{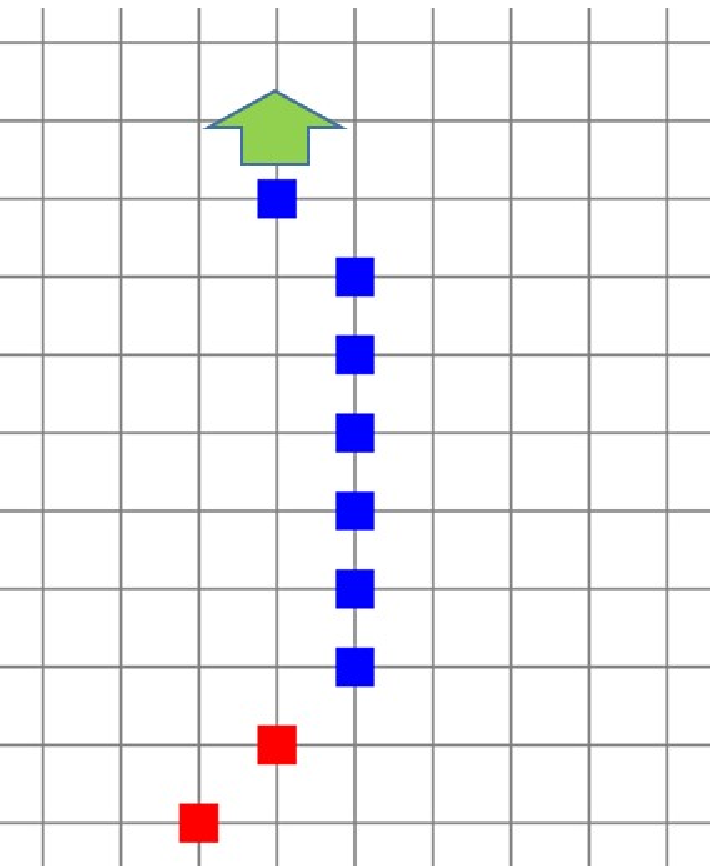}}
        \hspace{10pt}
		\subfloat[Rule 23]{\includegraphics[scale=0.35]{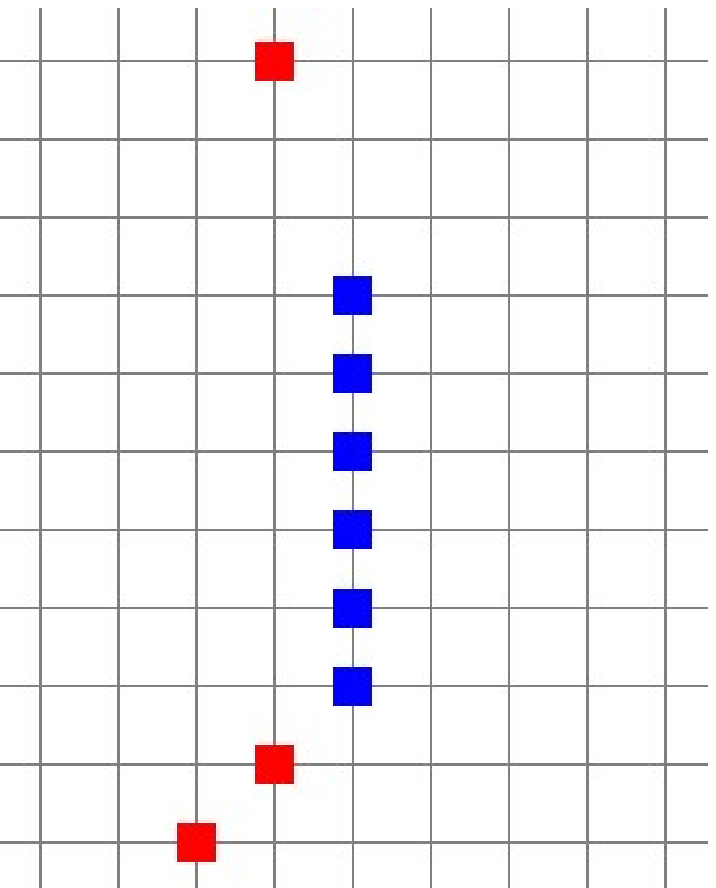}}
	\end{center}
    \caption{Examples of movements by Rules 20 to 23}
    \label{fig:Rule20to23}
\end{figure}

Remind that in configuration $C_D$, which is the final configuration of the first phase,
all robots with color $\CB$ are lined up with no gaps.
Therefore, there exists only one robot (robot $\alpha$) which cannot observe any other robot 
in the negative direction of Y-axis, and robot $\alpha$ moves to the negative direction of X-axis 
by Rule 16 (Figure \ref{fig:Rule16to19}(a)).
By this movement, all other robots can observe a robot with $\CB$ in their third quadrant,
and the robots changes their colors into $\CA$ by Rule 17 (Figure \ref{fig:Rule16to19}(b)).
If robot $\alpha$ observes no robot with color $\CB$, 
it moves again to the negative direction of X-axis by Rule 18 (Figure \ref{fig:Rule16to19}(c)).
After that, the robot with the smallest y-coordinate among the robots with color $\CA$ moves to 
the negative direction of X-axis by Rule 19 (Figure \ref{fig:Rule16to19}(a)), and changes its color into $\CB$ by Rule 20 (Figure \ref{fig:Rule20to23}(a)).
The robot with the largest y-coordinate (robot $\beta$) moves to the negative direction of X-axis by Rule 21 
(Figure \ref{fig:Rule20to23}(b)). 
It is worthwhile to mention that robot $\beta$ can observe both robots with color $\CB$
because the number of the robots $n$ is greater than 3.
Furthermore, robot $\beta$ can observe all other robots after the execution of Rule 21, 
this means that robot $\beta$ can count the number of robots $n$ even it moves to the positive direction of Y-axis.
Robot $\beta$ can repeatedly move to the positive direction of Y-axis
until it reaches (1,$m$-1) (assuming that robot $\alpha$ is on the origin) by Rule 22 (Figure \ref{fig:Rule20to23}(c)).
When robot $\beta$ reaches (1,$m$-1), it changes the color of its light into $\CB$ by Rule 23 (Figure \ref{fig:Rule20to23}(d)).
As a result, the configuration satisfies configuration $C_E$, and the second phase is completed.

\subsection{Phase 3: The Complete Visibility Achievement Phase}
Algorithm \ref{algo_phase3} shows the third phase of 
the proposed algorithm, 
called \emph{the Complete Visibility Achievement Phase}, 
which consists of 3 rules, from Rule 24 to Rule 26.

\begin{algorithm}[tbhp]
    \caption{Complete Visibility Achievement Phase}
    \begin{algorithmic}
\STATE {\it /* From configuration $C_E$ to configuration $C_F$ */}

       \STATE {\bf Rule24:} $\myC = \CA \land (x \leq -1, y \geq 1, \CB) \geq 1 \land (x \leq -1, y \leq -1, \CB) \geq 1 \land
                                   (x \leq -1, -\infty \leq y \leq \infty, \CA) = 0 \land \LookCS = 1 \land
                                   \MyLocation \neq (\yi^2 \bmod m,\yi) \to (x^+,\bot)$
       \STATE {\bf Rule25:} $\myC = \CA \land (x \leq -1, y \geq 1, \CB) \geq 1 \land (x \leq -1, y \leq -1, \CB) \geq 1 \land
                                   (x \leq -1, -\infty \leq y \leq \infty, \CA) = 0 \land \LookCS = 0 \to (x^+,\bot)$

       \STATE {\bf Rule26:} $\myC = \CA \land (x \leq -1, y \geq 1, \CB) \geq 1 \land (x \leq -1, y \leq -1, \CB) \geq 1 \land
                                   (x \leq -1, -\infty \leq y \leq \infty, \CA) = 0 \land \LookCS = 1 \land
                                   \MyLocation = (\yi^2 \bmod m,\yi) \to (\bot,\CB)$
    \end{algorithmic}
  \label{algo_phase3}
\end{algorithm}

We define the final configuration $C_F$ as the following.
Note that the complete visibility problem is solved when configuration $C_F$ is achieved 
because any three robots in $C_F$ are not collinear.

\noindent
{\bf Configuration $C_F$}: 
Let a robot with color $\CB$ with the smallest y-coordinate be the origin, 
  all robots have light with color $\CB$ and 
  are located at ($i^2 \bmod m, i$) for distinct $i$ ($i \in \mathbb{N}_0$), 
  where $m$ is the smallest prime number greater than or equal to the number of robots $n$.

As the result of the second phase, {\em Complete Visibility Achievement Phase}, 
the coordinate system is defined by the three robots with color $\CB$, 
which are located at (0,0), (1,1), and (1,$m$-1) respectively 
assuming that the point occupied by the robot with the smallest y-coordinate (robot $\alpha$) is the origin.
All other robots than the three reference robots can observes robots both in the second and third quadrants 
only during the third phase. 
In Phase 3, we use two functions, $\LookCS$ and $\MyLocation$:
$\LookCS$ returns 1 if a robot can observe all three reference robots with color $\CB$, 
and $\MyLocation$ returns its current coordinate based on the generated coordinate system by the three reference robots.
Note that all robots have distinct y-coordinates since the end of the first phase thus we use 
each robot's y-coordinate as distinct $i \in \mathbb{N}_0$.
In the proposed algorithm, $\yi$ represents the robot's y-coordinate based on the generated coordinate system 
(it is obvious that each robot knows $\yi$ only when $\LookCS = 1$).

\begin{figure}[tbhp]
 \begin{minipage}{0.5\hsize}
  \begin{center}
   \includegraphics[scale=0.35]{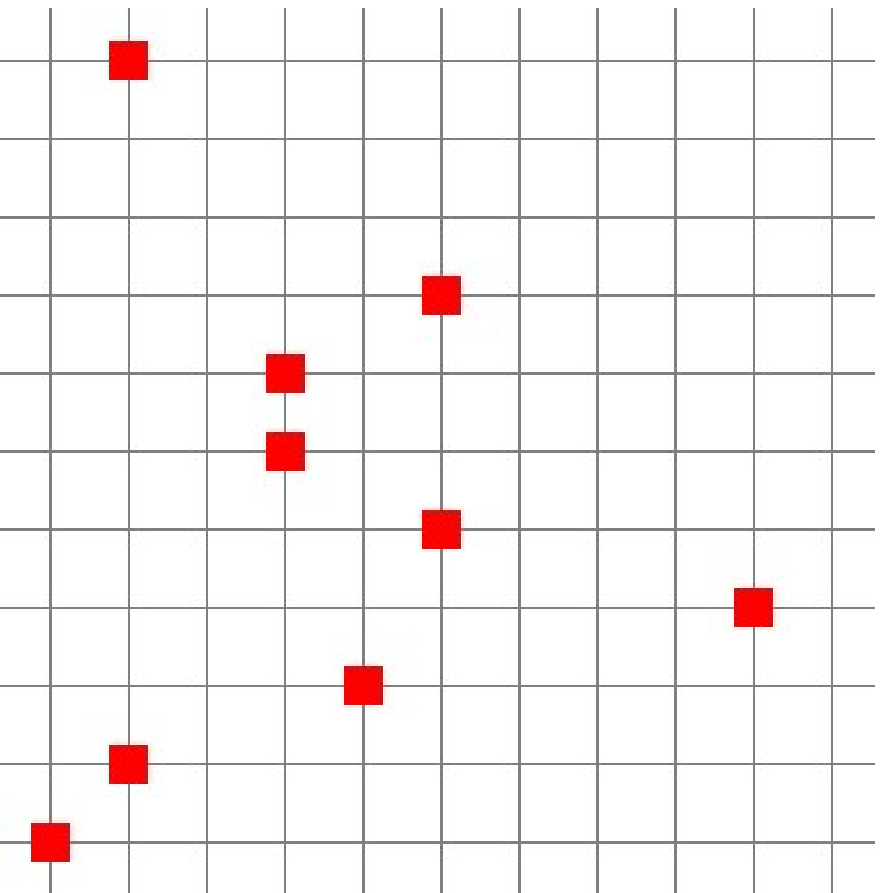}
  \end{center}
  \caption{Example of $C_F$}
  \label{fig:confF}
 \end{minipage}
 \begin{minipage}{0.5\hsize}
  \begin{center}
   \includegraphics[scale=0.35]{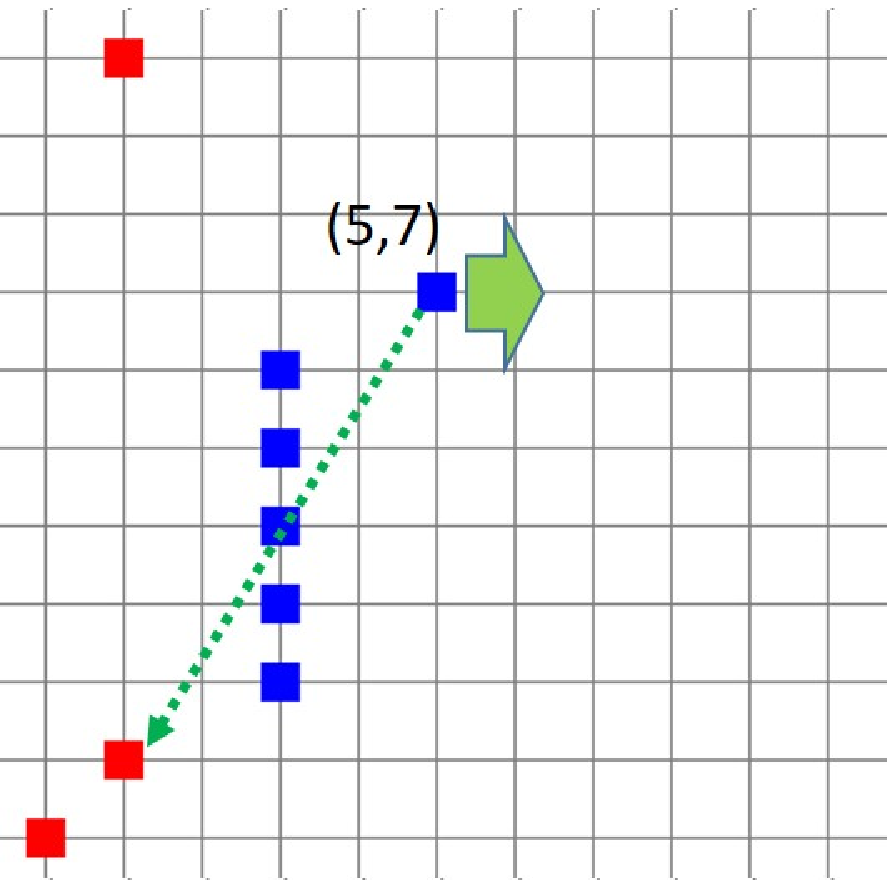}
  \end{center}
  \caption{Problem by obstruction}
  \label{fig:confFerror}
 \end{minipage}
\end{figure}

The main strategy of this phase is that 
each robot with color $\CA$ repeatedly moves to the positive direction of X-axis 
until it reaches ($\yi^2 \bmod m$,$\yi$) where $\yi$ is its y-coordinate based on the defined coordinate system, 
and changes its color into $\CB$ when it reaches ($\yi^2 \bmod m$,$\yi$).
Hence, when all robots change their color into $\CB$, the complete visibility is achieved,
\ie configuration $C_F$ is achieved (refer to Figure \ref{fig:confF}).

However, every robot with color $\CB$ moves asynchronously, 
there can be exist a robot cannot observe some reference robots caused by the obstruction by the other robots 
(Figure \ref{fig:confFerror}).
Therefore, we consider the pseudo-synchronous movement of the robots with color $\CA$:
each robot with color $\CA$ can move only when no other robot with color $\CA$ exists in the negative side of X-axis.
This implies that each robot with color $\CA$ has to wait 
until all other robots with color $\CA$ reach the point on the same Y-axis.
The proposed algorithm uses 3 rules to achieve the third phase
(refer to Algorithm \ref{algo_phase3}).

\begin{figure}[tbhp]
	\begin{center}
		\subfloat[Rule 24]{\includegraphics[scale=0.35]{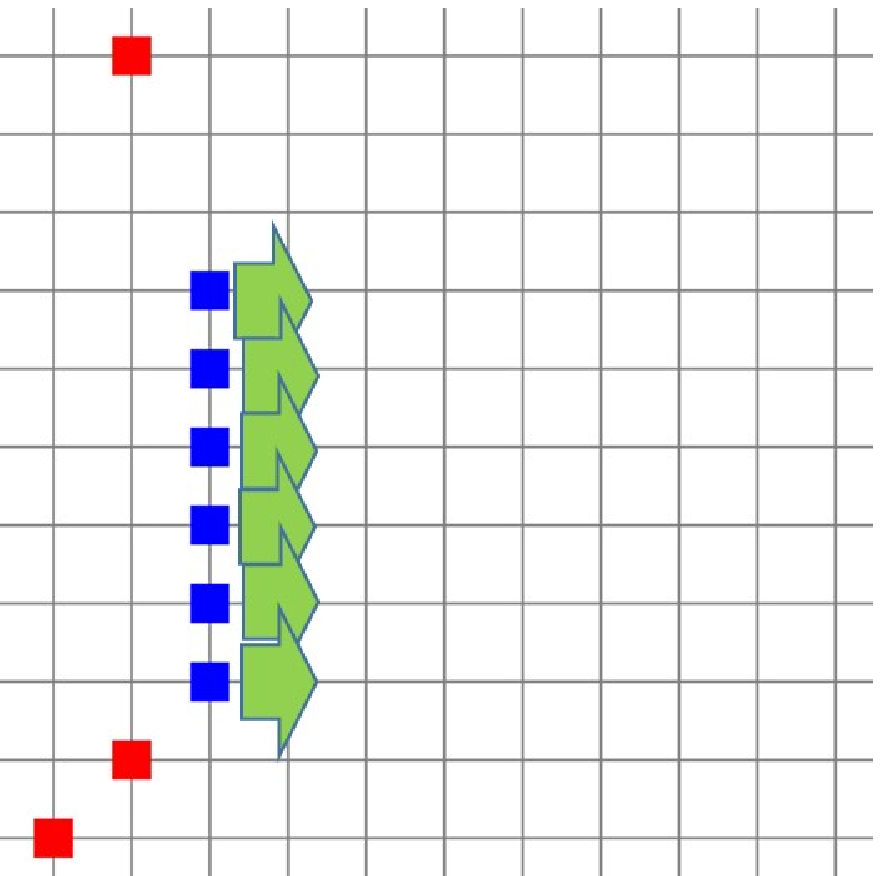}}
        \hspace{15pt}
		\subfloat[Rule 25]{\includegraphics[scale=0.35]{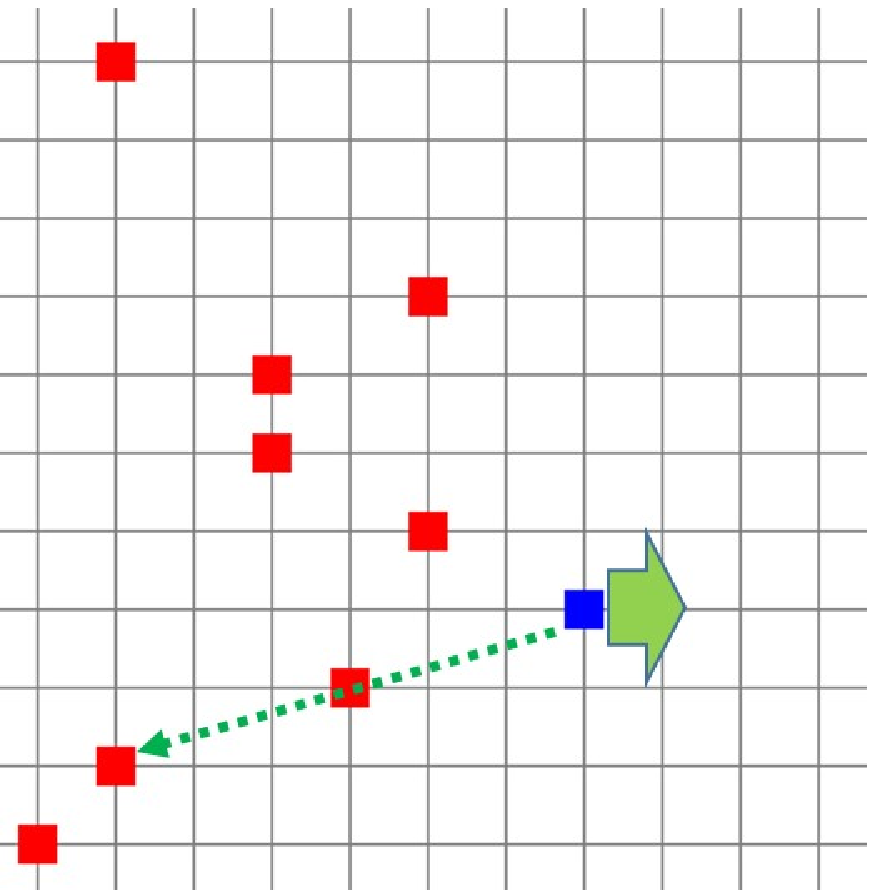}}
        \hspace{15pt}
		\subfloat[Rule 26]{\includegraphics[scale=0.35]{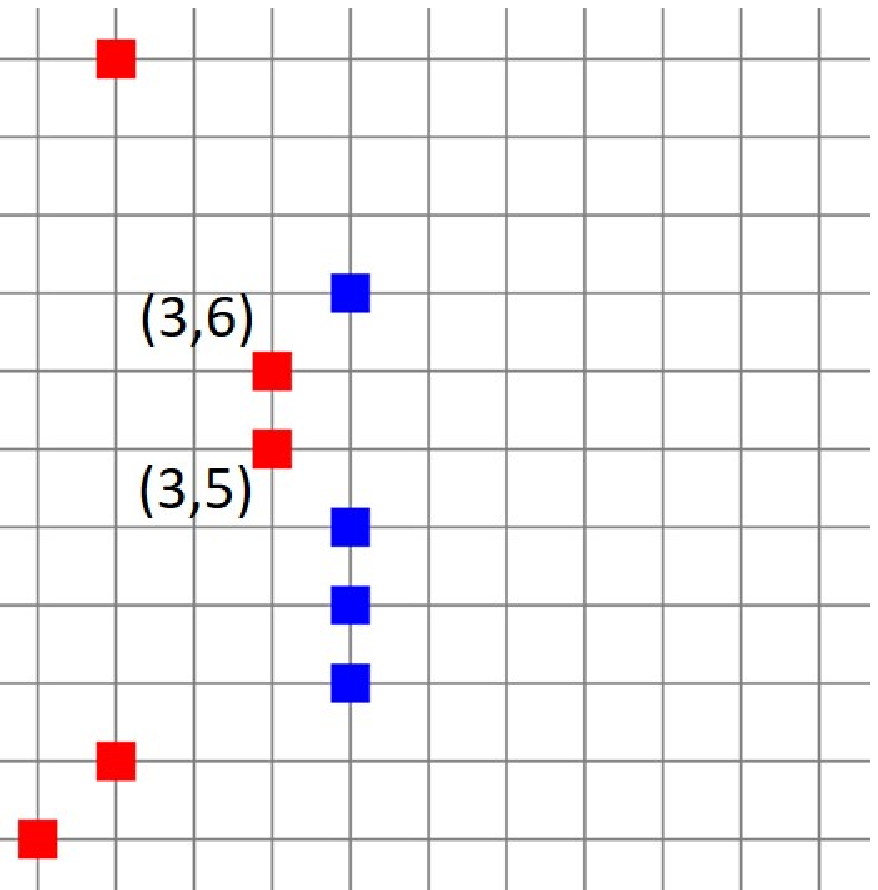}}
	\end{center}
    \caption{Examples of movement by Rules 24 to 26}
    \label{fig:Rule242526}
\end{figure}

Basically, each robot with color $\CA$ checks whether its current position is ($\yi^2 \bmod m$,$\yi$) or not:
if it is located on ($\yi^2 \bmod m$,$\yi$), it changes its color into $\CB$ and stops by Rule 26 (Figure \ref{fig:Rule242526}(c)),
otherwise it moves to the positive direction of X-axis 
unless there exist a robot with color $\CA$ on the negative side of X-axis by Rule 24 (Figure \ref{fig:Rule242526}(a)).
If a robot with color $\CA$ cannot observe the three reference robots,
it moves to the positive direction of X-axis without regard to its current position by Rule 25 (Figure \ref{fig:Rule242526}(b)):
this is because that if the robot is located on ($\yi^2 \bmod m$,$\yi$), 
it can observe all the reference robots.

\subsection{Analysis of the Algorithm}
\begin{theorem}
\label{correct}
The proposed algorithm guarantees to solve the complete visibility problem
using only two colors, $\CC = \{ \CA, \CB \}$, under an asynchronous scheduler
in $\mathcal{O}(D+n)$ asynchronous rounds and within $\mathcal{O}(n^2)$ area
when an arbitrary set of robots with color $\CA$ is given.
\end{theorem}

We omit the proof of the correctness of the proposed algorithm,
but here we briefly give a proof sketch of the time complexity of the proposed algorithm.
We analyze the time complexity using an asynchronous round that is defined as the shortest sequence of an execution 
where each robot performs its three operations at least once.

The worst case of Phase 1 is a configuration so that all robots except two robots, 
$\alpha$ and $\beta$, are lined up along X-axis:
only one robot may move in the first round after this configuration under an adversary scheduler 
because all robots located on $y_i$ (where $i \leq -2$) cannot move even they are activated by the scheduler.
However, two robots can move in the next round regardless of the sequence of activation, 
as the same manner, $i$ robots can move in the $i$-th round.
the head (leading) robot may stop only one round to wait until robot $\beta$ moves, 
which causes all robots to stop once: 
this implies that the tail robot may stop at most $\BigO(n)$ rounds.
As a result, from any initial configuration, 
all robots are lined up along X-axis in $\BigO(D)$ rounds 
(where $D$ is the diameter of the initial configuration),
the robot farthest away from the target line, $y_0$, begins to move after $\BigO(n)$ rounds,
and it moves $\BigO(n)$ times but it may wait at most $\BigO(n)$ times.
Hence, Phase 1 requires $\BigO(D+n)$ rounds.

In Phase 2, two robots (other than robot $\beta$) among the three reference robots move the constant number of rounds
and one robot (robot $\beta$) moves $\BigO(m)$ times.
Thus Phase 2 converges in $\BigO(n)$ rounds by \emph{Bertrand's postulate}; 
for any integer $n > 3$, there always exists at least one prime number $p$ with $n < p < 2n-2$.

In Phase 3, every robot which is not located on the goal position 
can move to the positive direction of X-axis in every round, at most $\BigO(m)$ times.
As the same reason of Phase 2, Phase 3 also converges in $\BigO(n)$ rounds.
Hence the proposed algorithm terminates in $\BigO(D+n) + \BigO(n) + \BigO(n) = \BigO(D+n)$ rounds.

It is obvious that the spatial complexity of the proposed algorithm is 
$\mathcal{O}(n^2)$ due to the deployment strategy: $(i^2 \bmod m, i)$ for distinct $i < n$ (note that $n < m$).


\section{Conclusion}
In this paper, we consider the complete visibility problem 
which is a problem to relocate all robots on grid plane to observe each other, \ie no three robots are collinear.
We proposed an (asynchronous) algorithm for oblivious robots 
to solve the complete visibility problem in $\BigO(D+n)$ asynchronous rounds
under the following assumptions:
the robots (1) agree on both axes, X-axis and Y-axis, but do not agree on the position of the origin, 
(2) have unlimited visibility range, 
(3) are not transparent,
(4) do not know the number of robots, 
and (5) have the light with only 2 colors.


Compared with the previous studies, the proposed algorithm uses the fewest number of colors 
to achieve complete visibility, even we assume the stronger geometrical agreement.
In many previous works for the luminous model, \ie robots with lights, 
the necessary number of colors is one of the important issue.
As the main contribution of this work, 
we show that the necessary number of colors to solve the complete visibility problem on grid plane 
can reduce to 2 if robots have the common sense of direction, \ie they agree on the both axes.

As a future work, 
we consider the weaker assumptions of the geometrical agreement, \eg only one axis or chirality.
To clarify the possibility of the algorithm by the robots without light, \ie fully oblivious robots, 
is another future work: to propose an algorithm or to provide the impossibility result.

\section*{Acknowledgements}
This work was supported in part by JSPS KAKENHI Grant Numbers 19K11823, 20K11685, 20KK0232,
and 21K11748.

\bibliographystyle{unsrtnat}
\bibliography{arXiv_CV}







\end{document}